\documentclass[preprint,12pt,review]{elsarticle}

\usepackage{amssymb}
\usepackage{amsthm}
\usepackage{amsmath}
\usepackage{mathrsfs}
\usepackage{graphicx}
\usepackage{epstopdf}
\usepackage{float}
\usepackage{caption}
\usepackage{subcaption}
\usepackage{bm}
\usepackage{amssymb} 
\usepackage{extarrows} 
\usepackage{bbm}
\usepackage{mathrsfs}
\usepackage{cleveref}
\usepackage{soul}
\usepackage{siunitx}
\usepackage{tabu} 
\usepackage{longtable}
\usepackage{xcolor} 
\usepackage{comment} 
\usepackage[margin=1.75cm]{geometry}
\biboptions{sort&compress} 
\usepackage{upgreek} 

\usepackage{booktabs}
\usepackage{amsmath,bm}
\usepackage[title]{appendix}
\usepackage[makeroom]{cancel}
\usepackage{tikz}
\usepackage{mathtools}
\journal{International Journal of Fatigue}

\makeatletter
\def\@author#1{\g@addto@macro\elsauthors{\normalsize%
    \def\baselinestretch{1}%
    \upshape\authorsep#1\unskip\textsuperscript{%
      \ifx\@fnmark\@empty\else\unskip\sep\@fnmark\let\sep=,\fi
      \ifx\@corref\@empty\else\unskip\sep\@corref\let\sep=,\fi
      }%
    \def\authorsep{\unskip,\space}%
    \global\let\@fnmark\@empty
    \global\let\@corref\@empty  
    \global\let\sep\@empty}%
    \@eadauthor={#1}
}
\makeatother

\begin{document}

\begin{frontmatter}



\title{A phase field model for high-cycle fatigue: total-life analysis}


\author{Alireza Golahmar\fnref{DTU,IC}}

\author{Christian F. Niordson\fnref{DTU}}

\author{Emilio Mart\'{\i}nez-Pa\~neda\corref{cor1}\fnref{IC}}
\ead{e.martinez-paneda@imperial.ac.uk}

\address[DTU]{Department of Mechanical Engineering, Technical University of Denmark, DK-2800 Kgs. Lyngby, Denmark}

\address[IC]{Department of Civil and Environmental Engineering, Imperial College London, London SW7 2AZ, UK}

\cortext[cor1]{Corresponding author.}

\begin{abstract}
We present a generalised phase field formulation for predicting high-cycle fatigue in metals. Different fatigue degradation functions are presented, together with new damage accumulation strategies, to account for (i) a typical S-N curve slope, (ii) the fatigue endurance limit, and (iii) the mean stress effect. The numerical implementation exploits an efficient quasi-Newton monolithic solution strategy and \emph{Virtual} S-N curves are computed for both smooth and notched samples. The comparison with experiments reveals that the model can accurately predict fatigue lives and endurance limits, as well as naturally capture the influence of the stress concentration factor and the load ratio.\\
\end{abstract}


\end{frontmatter}

\section{Introduction}
The fracture of materials subjected to fatigue loading is arguably the main failure mechanism of engineering components, accounting for (up to) 90$\%$ of all structural failures \cite{stephens2000}. Due to its complexity, the development of numerical methods capable of predicting fatigue cracking is of great utility and has been a prominent research field for several decades. Generally, the evolution of fatigue damage can be divided into two stages: (i) crack nucleation and (ii) crack growth. In the initiation stage, permanent microscopic degradation phenomena such as micro-voids and, subsequently, micro-cracks are formed in the material. These micro-cracks start growing and eventually coalesce, leading to the formation of dominant fatigue (macro-) cracks. One or more of those macro-cracks will then propagate, first in a stable manner, and finally unstably leading to the complete failure of the component.

Fatigue design is commonly based on classical empirical methods which involve data fitting of a large number of experimental tests \cite{Suresh1998}. Such methods estimate the fatigue life as a function of the cyclic stress (or strain) range, where the fatigue life is defined as the number of cycles ($N_f$) or reversals ($2N_f$) to failure. A pioneering work in this area is that of Wöhler \cite{Wohler1870}, which is commonly referred to as the stress-life or S-N curve approach. In general, fatigue life analyses are divided into two limiting cases. One is denoted as high-cycle fatigue (HCF), a regime where the material is exposed to low cyclic stress amplitudes, behaving mainly in an elastic manner and requiring a large number of cycles to fail (often up to $10^6$ cycles). This approach has become popular in applications involving low-amplitude cyclic stresses such as offshore wind structures exposed to alternating mechanical loads caused by the wind and sea waves. A second scenario is that where the applied stresses are large enough to cause plastic deformations and thus a much lower number of cycles are needed to see failure; $10^4$ cycles or fewer, a regime referred to as low-cycle fatigue (LCF). Due to their empirical nature, stress-life methods have limited applicability and can be barely generalised to arbitrary materials, geometries and loading conditions.

Variational phase field fracture models can provide a mechanistic computational framework to predict low- and high- cycle fatigue, overcoming the challenges of empirical methods. The model is based upon Griffith’s thermodynamical framework \cite{Griffith1920}, whereby a crack would grow if the energy released by the solid exceeds its critical value, the material toughness. Francfort and Marigo \cite{Francfort1998} presented a variational formulation for Griffith’s energy balance, and Bourdin et al. \cite{Bourdin2008} introduced a scalar phase field variable to regularise the resulting functional and obtain computational predictions of crack evolution as an exchange between stored and fracture energy. Since its early development, the phase field fracture method has been gaining increasing attention and its use has been extended to numerous applications, including ductile damage \cite{Ambati2015b,Borden2016,Isfandbod2021}, dynamic fracture \cite{Borden2012,Geelen2019,Molnar2020}, composites delamination \cite{Alessi2019,Mandal2020,Quintanas-Corominas2020}, fracture of functionally graded materials \cite{Hirshikesh2019,Kumar2021}, and hydrogen-assisted cracking \cite{Martinez-Paneda2018,Duda2018,Wu2020a}, among many others; see Refs. \cite{Wu2020b,Kristensen2021} for an overview.

Recently, efforts have been made to incorporate fatigue damage into variational phase field fracture methods. Lo et al. \cite{Lo2019} introduced a viscous term into the standard phase field model for brittle fracture, combined with a modified $J$-integral, to generate Paris-law type fatigue crack growth behaviour. More commonly, an additional variable describing the fatigue history is introduced. This variable has been defined either as a dissipative term to the microforce balance of the phase field \cite{Boldrini2016,Loew2020,Schreiber2020}, to effectively reduce crack growth resistance, or as a fatigue degradation function that reduces the material toughness \cite{Alessi2018,Carrara2020,Seiler2020,Simoes2021,Simoes2022,Ai2022}. 
Accordingly, an additional equation is introduced to describe the evolution/accumulation of the fatigue history variable. Boldrini et al. \cite{Boldrini2016} derived this additional equation from thermodynamic principles while Loew et al. \cite{Loew2020} proposed an equation based on micro-crack growth. Seiler et al. \cite{Seiler2020} applied a local strain approach to empirically incorporate plasticity via Neuber’s rule, while Schreiber et al. \cite{Schreiber2020} employed Miner’s rule to govern the evolution of fatigue damage. Alessi and co-workers \cite{Alessi2018} proposed describing the evolution of fatigue damage as a function of the accumulated strain during the loading stage of each cycle. Following \cite{Alessi2018}, the authors of \cite{Carrara2020,Hasan2021,Golahmar2022,Simoes2021,Seles2021,Ulloa2021,Khalil2022} accumulated the tensile (non-compressive) parts of the strain energy density (elastoplastic energy density in \cite{Ulloa2021,Khalil2022}) only during the loading (unloading in \cite{Seles2021}) stages.

In this work, we present a generalised formulation for modelling the fatigue behavior of metallic materials. We restrict our attention to high-cycle fatigue (HCF) analysis and build our formulation upon the variational phase field approach for fatigue proposed by Alessi et al. \cite{Alessi2018} and Carrara et al. \cite{Carrara2020}. New accumulation strategies for the evolution of fatigue damage are proposed, so as to capture the typical S-N curve slope, the fatigue endurance limit and the mean stress effect (load/stress ratio). The framework encompasses the two most widely used phase field fracture models, so-called \texttt{AT1} \cite{Pham2011} and \texttt{AT2} \cite{Bourdin2008}. Importantly, the numerical implementation makes use of a quasi-Newton monolithic solution scheme \cite{Wu2020c,Kristensen2020}, which is essential to minimise the cost of cycle-by-cycle fatigue simulations. Moreover, the new accumulation strategy presented further accelerates computations since, as described below, it enables solving the coupled system of iterations only once per loading cycle.

The theoretical elements of the new generalised phase field fatigue framework presented are first described in Section \ref{Sec:Theory}. Then, in Section \ref{Sec:Numerical}, details of the numerical implementation are provided. The results obtained are given in Section \ref{Sec:Results}. Several boundary value problems have been addressed to investigate the performance of the proposed modelling framework. First, the response of a homogeneous bar under uniaxial cyclic/monotonic loading is thoroughly studied to showcase the influence of the different material/model parameters introduced. In addition, the failure of a notched cylindrical bar is predicted for different load ratios and notch radii, and predictions are compared with fatigue experiments (S-N curves) on two types of steel; AISI 4340 and 300M. Finally, the manuscript ends with concluding remarks in Section \ref{Sec:Conclusions}.

\section{A phase field model for fatigue damage}

\label{Sec:Theory}
The formulation presented in this section refers to the response of an elastic solid body occupying the volume $\Omega \subset \mathbb{R}^{\delta} \; (\delta \in [1,2,3])$ having the external surface $\partial \Omega \subset \mathbb{R}^{\delta -1}$ with the outward unit normal $\mathbf{n}$. We first define the field variables of the model (Section \ref{Sec:Kinematics}), then derive the balance of forces using the principle of virtual power (Section \ref{Sec:VWP}), proceed to formulate the local free-energy imbalance under isothermal conditions (Section \ref{sec:EnergyImbalance}), and finally particularise our theory to suitable constitutive choices for the deformation, fracture and fatigue behaviour of the solid (Section \ref{sec:ConstitutiveTheory}).

\subsection{Field variables and kinematics}
\label{Sec:Kinematics}
The primary field variables are the displacement field vector $\textbf{u}$ and the damage phase field $\phi$. Assuming small deformations, the strain tensor $\bm{\varepsilon}$ is given by
\begin{equation}
    \bm{\varepsilon}=\dfrac{1}{2}\left(\nabla^{\mathsf{T}}\mathbf{u}+\nabla\mathbf{u}\right)
\end{equation}

The nucleation and growth of fatigue cracks are described by using a smooth continuous scalar phase field $\phi \in [0; 1]$. The use of an auxiliary phase field variable to implicitly track interfaces has proven to be a very compelling computational approach for numerous interfacial problems, such as microstructural evolution \cite{Provatas2011} and metallic corrosion \cite{Cui2021}. In the context of fracture mechanics, the phase field variable resembles a damage variable; it must grow monotonically $\dot{\phi}\geq0$ and describes the degree of damage, with $\phi=1$ denoting a crack and $\phi=0$ corresponding to intact material points. Since $\phi$ is smooth and continuous, discrete cracks are represented in a diffuse fashion, with the smearing of cracks being controlled by a phase field length scale $\ell$. The aim of this diffuse representation is to introduce, over a discontinuous surface $\Gamma$, the following approximation of the fracture energy \cite{Bourdin2008}:
\begin{equation}
    \Psi^{s}=\int_{\Gamma} G_{c}\, \mathrm{d}S \approx \int_{\Omega} G_{c} \gamma_{\ell}(\phi,\nabla\phi)\, \mathrm{d}V\,, \hspace{10mm} \text{for}\hspace{5mm} \ell \rightarrow 0^+\,,
\end{equation}
\noindent where $\gamma_{\ell}$ is the so-called crack surface density functional and $G_{c}$ denotes the critical Griffith-type energy release rate, or material toughness. We extend this rate-independent description of fracture to accommodate time and history dependent problems. Thus, for a cumulative history variable $\bar{\alpha}$, which fulfils $\dot{\bar{\alpha}}\geq0$ for a current time $\tau$, and a fatigue degradation function $f(\bar{\alpha})$, the fracture energy can be re-formulated as follows
\begin{equation}
    \Psi^{s}=\int_{0}^{t}\int_{\Omega} f(\bar{\alpha}(\tau)) \, G_{c} \, \dot{\gamma}_{\ell}(\phi,\nabla\phi)\, \mathrm{d}V \mathrm{d}\tau
    \label{eq:Pis^s_fatigue}
\end{equation}
\subsection{Principle of virtual power. Balance of forces}
\label{Sec:VWP}

The balance equations for the coupled problem are now derived using the principle of virtual power. With respect to the displacement $\textbf{u}$, the external surface of the body is decomposed into a part $\partial \Omega_u$, where the displacement is prescribed by Dirichlet-type boundary conditions, and a part $\partial \Omega_h$, where the traction $\mathbf{h}$ is prescribed by Neumann-type boundary conditions. A body force field per unit volume $\mathbf{b}$ can also be prescribed. With respect to the phase field $\phi$, a Dirichlet-type boundary condition can be prescribed at $\Gamma$, a given crack surface inside the solid body. Additionally, a phase field fracture microtraction $f$ can be prescribed on $\partial\Omega_{f}$. Accordingly, the external and internal virtual powers read
\begin{equation}
\label{eq:VWP}
\begin{split}
   \dot{\mathcal{W}}_{\text{ext}} &=\int_{\partial \Omega}\Big\{\mathbf{h} \cdot \dot{\mathbf{u}}+f \dot{\phi}\Big\} \,\mathrm{d}S + \int_{\Omega} \mathbf{b} \cdot \dot{\mathbf{u}}\, \mathrm{d}V\\[3mm]
    \dot{\mathcal{W}}_{\text{int}} &= \int_{\Omega}\left\{\bm{\sigma}: \nabla \dot{\mathbf{u}}+\omega \dot{\phi}+\bm{\upxi} \cdot \nabla \dot{\phi}\right\} \, \mathrm{d} V
\end{split}
\end{equation}
\noindent where $\bm{\sigma}$ is the Cauchy stress tensor work conjugate to the elastic strains $\bm{\varepsilon}$, and $\omega$ and $\bm{\upxi}$ are the microstress quantities work conjugate to the phase field $\phi$ and its gradient $\nabla \phi$, respectively.  Eq. (\ref{eq:VWP}) must hold for an arbitrary domain $\Omega$ and for any kinematically admissible variations of the virtual quantities.  Thus, by application of the Gauss divergence theorem and the fundamental lemma of calculus of variations, the local force balances (in $\Omega$) are given by
\begin{equation}
\begin{aligned}
\nabla \cdot \bm{\sigma}+\mathbf{b}=\mathbf{0} & \\[3mm]
\nabla \cdot \bm{\upxi}-\omega=0 & 
\end{aligned}
\label{eq:SF3}
\end{equation}
along with the following natural boundary conditions (on $\partial\Omega$)
\begin{equation}
\begin{aligned} \hspace{13mm}
\mathbf{h}=\bm{\sigma} \cdot \mathbf{n} \\[3mm]
f =\bm{\upxi} \cdot \mathbf{n} & 
\end{aligned}
\label{eq:NTBC3}
\end{equation}
\subsection{Free-energy imbalance}
\label{sec:EnergyImbalance}
The first and second law of thermodynamics can be expressed through the Helmholtz free energy per unit volume $\psi\left(\bm{\varepsilon},\phi,\nabla \phi \right)$ and the external work $\mathcal{W}_{\text{ext}}$,
\begin{equation} \label{eq:CDI3}
    \int_{\Omega} \dot{\psi} \,\mathrm{d} V-\int_{\partial \Omega} \dot{\mathcal{W}}_{\text{ext}} \, \mathrm{d} S \leq 0
\end{equation}
which is generally referred to as Clausius–Duhem inequality. Inserting Eqs. (\ref{eq:SF3})-(\ref{eq:NTBC3}) and applying the divergence theorem, the local free-energy inequality can be rewritten as
\begin{equation}
    \int_{\Omega} \dot{\psi}\, \mathrm{d} V-\int_{\Omega}\left\{\bm{\sigma}: \nabla \dot{\mathbf{u}}+\omega \dot{\phi}+\bm{\upxi} \cdot \nabla \dot{\phi}\right\} \mathrm{d} V \leq 0
\end{equation}
which must hold for any arbitrary volume and, thus, must also hold in a local fashion,
\begin{equation}
    \left(\bm{\sigma}-\frac{\partial \psi}{\partial \bm{\varepsilon}}\right): \dot{\bm{\varepsilon}}+
    \left(\omega-\frac{\partial \psi}{\partial \phi}\right) \dot{\phi}+
    \left(\upxi-\frac{\partial \psi}{\partial \nabla \phi}\right) \cdot \nabla \dot{\phi} \geq 0
    \label{eq:DC3}
\end{equation}
for which a free energy function $\psi$ is proposed as the sum of the elastic strain energy density of the
solid $\psi^e$ and the fracture surface energy density $\psi^s$, such that:
\begin{equation}
    \psi(\bm{\varepsilon}, \phi, \nabla \phi \,|\: \bar{\alpha})= \psi^e(\bm{\varepsilon},\phi) + \psi^s( \phi, \nabla \phi \,|\: \bar{\alpha})
    \label{eq:FreeEnergyDensity}
\end{equation}
\subsection{Constitutive theory}
\label{sec:ConstitutiveTheory}
Consistent with the free energy definition (\ref{eq:FreeEnergyDensity}), we proceed now to develop a constitutive theory that couples the deformation, fracture and fatigue behaviour of the solid.
\subsubsection{Elasticity}
The strain energy density $\psi^{e}$ is defined as a function of the elastic strains $\bm{\varepsilon}$, the isotropic linear elastic stiffness tensor $\bm{\mathcal{L}}_{0}$ and a phase field degradation function $g(\phi)$, to be defined. Hence,
\begin{equation}
    \psi^e(\bm{\varepsilon},\phi)=g(\phi)\psi^{e}_{0}(\bm{\varepsilon})\hspace{10mm}\text{with}\hspace{5mm}\psi^{e}_{0}(\bm{\varepsilon})=\dfrac{1}{2}\bm{\varepsilon}^{\mathsf{T}} : \bm{\mathcal{L}}_{0} : \bm{\varepsilon}
    \label{eq:ElasticEnergyDensity}
\end{equation}
where $\psi^{e}_{0}$ denotes the strain energy density for an undamaged isotropic solid. Accordingly, the Cauchy stress tensor $\bm{\sigma}$ can now be derived as
\begin{equation}
    \bm{\sigma}=\frac{\partial \psi}{\partial \bm{\varepsilon}}=g(\phi) \bm{\mathcal{L}}_{0}: \bm{\varepsilon},
    \label{eq:CauchyStress}
\end{equation}
emphasising how the phase field order parameter reduces the stiffness of the solid, as in continuum damage mechanics approaches.
\subsubsection{Fracture surface energy}
The surface energy density of a fractured solid $\psi^s$, in agreement with (\ref{eq:Pis^s_fatigue}), is defined as a function of the phase field damage $\phi$, its gradient $\nabla \phi$ and a fatigue degradation function $f(\bar{\alpha})$, to be defined,
\begin{equation}
    \psi^s( \phi, \nabla \phi \,|\: \bar{\alpha})= f(\bar{\alpha}) \, G_{c} \, {\gamma}_{\ell}(\phi,\nabla\phi)
    \label{eq:FractureEnergyDensity}
\end{equation}
in which the crack surface density functional $\gamma_\ell$ is expressed as
\begin{equation}
    \gamma_{\ell}(\phi,\nabla\phi)=\frac{1}{4 c_{w}}\left(\frac{w(\phi)}{\ell}+\ell|\nabla \phi|^{2}\right) \hspace{10mm}\text{with}\hspace{5mm}  c_{w}=\int_{0}^{1} \sqrt{w(\zeta)} \mathrm{d} \zeta
\end{equation}
where $w(\phi)$ is the geometric crack function, to be defined, and $c_w$ is a scaling constant.
\subsubsection{Strain energy decomposition}
\label{sec:StrainEnergyDecomposition}
To prevent the nucleation and growth of cracks under compression, the strain energy density can be decomposed into active (tensile) and inactive (compressive) parts,
\begin{equation}
    \psi^{e}\left(\bm{\varepsilon},\phi\right)=g(\phi)\,\psi^{+}_{0}(\bm{\varepsilon})+\psi^{-}_{0}(\bm{\varepsilon})
    \label{eq:Decomposition}
\end{equation}
\noindent where we follow the \textit{hybrid} formulation proposed by Ambati et al. \cite{Ambati2015a} in applying the decomposition only to the phase field evolution equation. Among the multiple decomposition splits proposed in the literature, the present work adopts the following choices:\\
\noindent i) \texttt{Spectral tension-compression} split by Miehe et al. \cite{Miehe2010a}:
\begin{equation}
    \psi_{0}^{\pm}(\bm{\varepsilon})=\frac{1}{2} \lambda\langle\operatorname{tr}(\bm{\varepsilon})\rangle_{\pm}^{2}+\mu \operatorname{tr}\left(\bm{\varepsilon}_{\pm}^{2}\right)\,, \hspace{1mm}\text{with}\hspace{2mm}  \bm{\varepsilon}_{\pm}=\sum_{i=1}^{3}\left\langle\varepsilon_{i}\right\rangle_{\pm} \bm{n}_{i} \otimes \mathbf{n}_{i}
    \label{eq:decMiehe}
\end{equation}
\noindent ii) \texttt{No-tension} split by Freddi et al. \cite{Freddi2010} (see also \cite{Lo2019} for 3D strain states):
\begin{equation}
    \psi_{0}^{\pm}(\bm{\varepsilon})=\frac{1}{2}\lambda\operatorname{tr}^2(\bm{\varepsilon}_\pm)+\mu \operatorname{tr}\left(\bm{\varepsilon}_{\pm}^{2}\right)\,, \hspace{1mm}\text{with}\hspace{2mm}  \bm{\varepsilon}_{\pm}=\text{sym}_{\pm}(\bm{\varepsilon})
    \label{eq:decFreddi}
\end{equation}
\noindent iii) \texttt{Volumetric-deviatoric} split by Amor et al. \cite{Amor2009}:
\begin{equation}
    \begin{array}{l}{\psi_{0}^{+}(\bm{\varepsilon})=\dfrac{1}{2} \left(\lambda+\frac{2}{3}\mu\right)\langle\operatorname{tr}(\bm{\varepsilon})\rangle_{+}^{2}+\mu\left(\bm{\varepsilon}^{\mathrm{dev}}: \bm{\varepsilon}^{\mathrm{dev}}\right)} \\[3mm] {\psi_{0}^{-}(\bm{\varepsilon})=\dfrac{1}{2} \left(\lambda+\frac{2}{3}\mu\right)\langle\operatorname{tr}(\bm{\varepsilon})\rangle_{-}^{2}}\end{array} \,, \hspace{1mm}\text{with}\hspace{2mm} \bm{\varepsilon}^{\mathrm{dev}}=\bm{\varepsilon}-\frac{1}{3} \operatorname{tr}(\bm{\varepsilon}) \mathbf{I}
    \label{eq:decAmor}
\end{equation}
\noindent where $\lambda$ and $\mu$ are the Lamé constants for an isotropic material and $\mathbf{I}$ is the identity matrix. Also, $\pm$ is the plus-minus sing and $\langle \square\rangle$ are the Macaulay brackets, such that $\langle \square\rangle_{\pm}:=\frac{1}{2}(\square \pm|\square|)$, and $\text{sym}_{\pm}(\bm{\varepsilon})$ is the positive/negative-definite symmetric part of the strain tensor. For the case of \texttt{Spectral} and \texttt{No-tension} splits, the infinitesimal strain tensor is given in terms of the principal strains $\left\{\varepsilon_{i}\right\}_{i=1}^{3}$ and principal strain directions $\left\{\mathbf{n}_{i}\right\}_{i=1}^{3}$.
\subsubsection{Irreversibility condition}
Damage is an irreversible process and, as a consequence, the phase field evolution law must fulfil the condition $\dot{\phi}\geq0$. To this end, we follow Miehe et al. \cite{Miehe2010b} and define a history variable field $\mathcal{H}$ for a current time $t$,
\begin{equation}
    \mathcal{H}=\max _{\tau \in[0, t]} \psi_{0}^+(\bm{\varepsilon}(\mathbf{x}, \tau))\,,
    \label{eq:H}
\end{equation}
\noindent which satisfies the Karush–Kuhn–Tucker (KKT) conditions for both loading and unloading stages,
\begin{equation}
    \psi_{0}^+ - \mathcal{H} \leq 0 \,\text{,} \hspace{10mm} \dot{\mathcal{H}} \geq 0 \,\text{,} \hspace{10mm} \dot{\mathcal{H}}(\psi_{0}^+-\mathcal{H})=0
    \centering
\end{equation}
\subsubsection{Phase field fracture}
We proceed to derive the phase field micro-stress quantities $\omega$ and $\bm{\upxi}$. First, considering, (\ref{eq:ElasticEnergyDensity}), (\ref{eq:FractureEnergyDensity}) and (\ref{eq:H}), the total free energy density of the solid (\ref{eq:FreeEnergyDensity}) renders
\begin{equation}
    \psi(\bm{\varepsilon}, \phi, \nabla \phi \,|\: \bar{\alpha})= g(\phi)\mathcal{H} + f(\bar{\alpha})\frac{G_{c}}{4 c_{w}}\left(\frac{w(\phi)}{\ell}+\ell|\nabla \phi|^{2}\right)
\end{equation}

Accordingly, the micro-stress variables $\omega$ and $\bm{\upxi}$ can readily be derived as
\begin{equation}
    \omega=\frac{\partial \psi}{\partial \phi}=g^{\prime}(\phi)\mathcal{H}+f(\bar{\alpha}) \frac{G_c}{4 c_{w} \ell} w^{\prime}(\phi) \,\,\,\,\,\,\,\,\,\,\,\,\,\,
    \bm{\upxi}=\frac{\partial \psi}{\partial \nabla \phi}=f(\bar{\alpha})\frac{G_c\ell}{2 c_{w}}  \nabla \phi
\end{equation}

Inserting these constitutive relations in the phase field local balance (\ref{eq:SF3}b) yields the strong form of the evolution of the crack phase field under fatigue loading,
\begin{equation}
    \frac{G_{c} f(\bar{\alpha})}{2 c_{w}}\left(\frac{w^{\prime}(\phi)}{2 \ell}-\ell \nabla^{2} \phi\right)-\frac{G_{c} \ell}{2 c_{w}} \nabla \phi \nabla f(\bar{\alpha})+g'(\phi)\mathcal{H}=0
    \label{eq:StrongFormPhi}
\end{equation}
\subsubsection{Degradation and dissipation functions}
First, we proceed to define the phase field degradation function $g(\phi)$, which governs the degradation of the stored elastic energy due to damage evolution, and must satisfy
\begin{equation}
    g(0)=1, \hspace{10mm} g(1)=0, \hspace{10mm} g'(\phi)\leq0 \hspace{7mm}\text{for}\hspace{4mm} 0\leq \phi \leq 1
\end{equation}
where the first two constraints are the limits for the fully intact and fully broken states while the last constraint ensures convergence of $\partial \psi/\partial \phi$ to a final value for the fully broken state. To this end, we adopt the widely used quadratic degradation function
\begin{equation}
    g(\phi)=(1-\phi)^2
    \label{eq:gphi}
\end{equation}

In addition, we define the damage dissipation function $w(\phi)$, which rules the energy dissipation due to the formation of a new crack, and must fulfil
\begin{equation}
    w(0)=0, \hspace{10mm} w(1)=w_{1}>0, \hspace{10mm} w^{\prime}(\phi) \geqslant0 \hspace{7mm}\text{for}\hspace{2mm} 0\leq \phi \leq 1
\end{equation}
for which we adopt what are arguably the two most widely used models in the literature, the so-called \texttt{AT1} \cite{Pham2011} and \texttt{AT2} \cite{Bourdin2008} phase field models. The specific choice $w(\phi)=\phi^2\,(c_w=1/2)$ renders the \texttt{AT2} model while $w(\phi)=\phi\, (c_w=2/3)$ corresponds to the \texttt{AT1} formulation. The latter introduces a purely elastic response prior to the onset of damage, unlike the \texttt{AT2} case, where $w^{\prime}(0)=0$. As a result, a damage driving force threshold $\mathcal{H}_\mathrm{min}$ should be defined for the \texttt{AT1} model, such that the history field (\ref{eq:H}) yields
\begin{equation}
    \mathcal{H}=\max \left\{\max _{\tau \in[0, t]} \psi_{0}^+(\bm{\varepsilon}(\mathbf{x}, \tau)),\: \mathcal{H}_\mathrm{min}\right\} \hspace{10mm}\text{with}\hspace{5mm} \mathcal{H}_\mathrm{min}=\frac{3G_c}{16\ell}
    \label{eq:H_AT1}
\end{equation}

Considering the homogeneous solution to (\ref{eq:StrongFormPhi}) provides further insight into the role of the phase field length scale $\ell$. Thus, in a 1D setting, for a sample with Young's modulus $E$, subjected to a uniaxial monotonic stress $\sigma=g \left( \phi \right) E \varepsilon$; the homogeneous solution for the stress reaches a maximum at the following critical strength and strain,
\begin{equation}
    \texttt{AT1}:\: \sigma_{c}=\sqrt{\dfrac{3 E G_{c}}{8\ell}},\hspace{2mm}\varepsilon_{c}=\sqrt{\frac{3 G_{c}}{8 \ell E}}\,,\hspace{10mm}\texttt{AT2}:\: \sigma_{c}=\dfrac{3}{16}\sqrt{\dfrac{E G_{c}}{3 \ell}},\hspace{2mm}\varepsilon_{c}=\sqrt{\frac{G_{c}}{3 \ell E}}
    \label{eq:sigmac}
\end{equation}
where $\ell$ is shown to be not only a regularising parameter but also a material property that defines the material strength. This enables phase field models to predict crack nucleation and naturally recover the transition flaw size effect \cite{Tanne2018,Kristensen2021}; i.e., capturing both toughness-dominated failures (for long cracks) and strength-dominated failures (short cracks). 
\subsubsection{Fatigue damage}
\label{Sec:FatigueDamage}
Phase field fatigue models have proven to be capable of capturing the nucleation and growth of fatigue cracks, and can naturally recover key features such as the Wöhler curve or Paris law behaviour \cite{Carrara2020}. However, existing models need to be enhanced to be able to capture behaviour frequently observed in experiments and widely embedded in fatigue design standards. In the context of total-life analyses, this includes the definition of suitable model/material parameters that enables capturing: (i) the slope of the S-N curve, (ii) the endurance limit of the material, and (iii) the load ratio effect. Thus, our work aims at developing a framework that can incorporate those additional modelling capabilities, and at showcasing the ability of this framework to reproduce experimental data and naturally capture the role of stress concentration factors (e.g., predicting the life of a notched component from a smooth S-N curve).

\begin{figure}[bht!]
    \centering
    \includegraphics[width=0.49\linewidth]{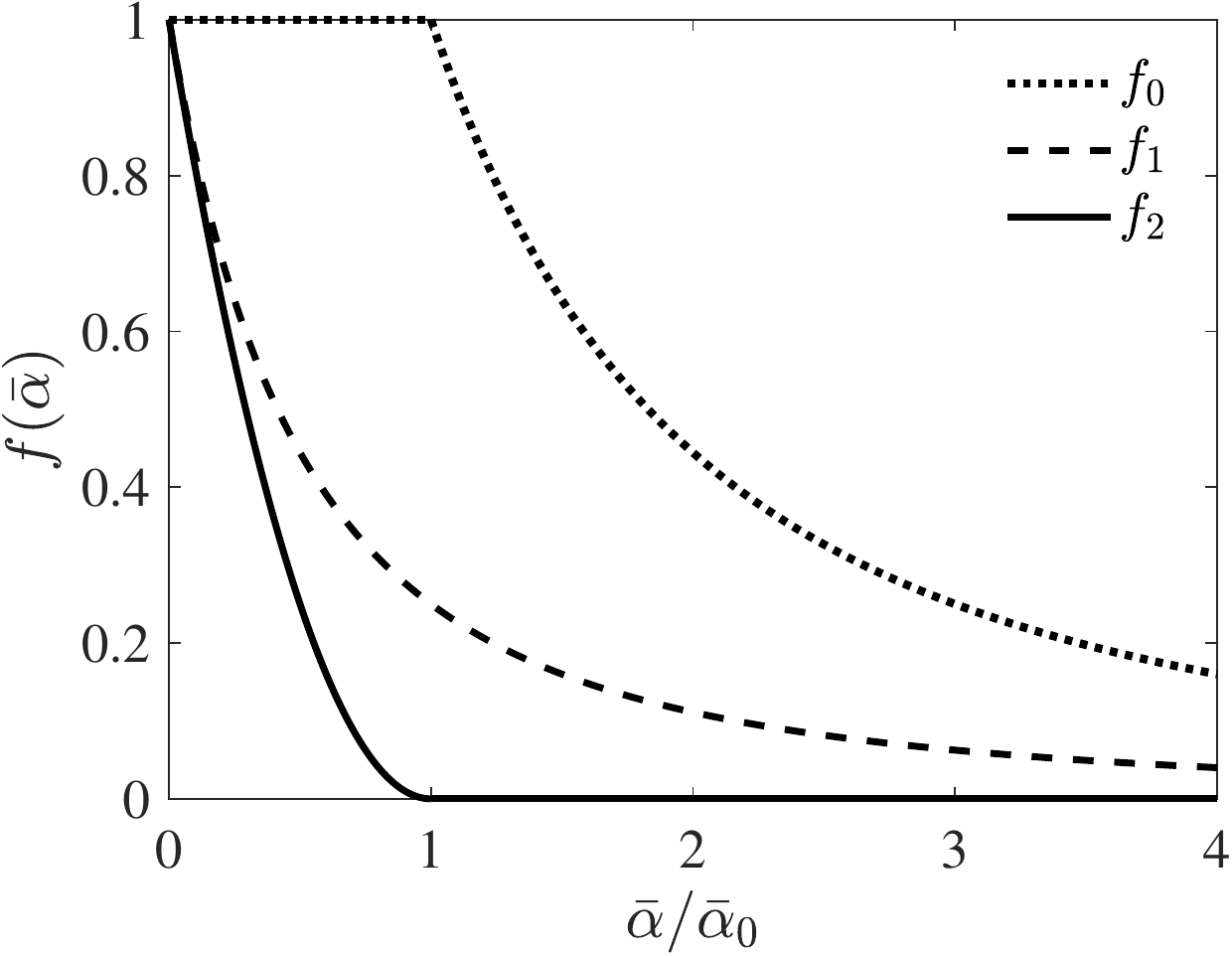}
    \caption{Evolution of the three fatigue degradation functions considered, see Eq. (\ref{eq:Fdegs}).}
    \label{fig:Fdegs}
\end{figure}

First, following \cite{Alessi2018}, the damage resulting from the application of cyclic loads is captured by introducing a fatigue degradation function $f(\bar{\alpha})$, which effectively degrades the material toughness as a function of the fatigue history experienced by the solid. The following fatigue degradation functions, proposed in the literature \cite{Carrara2020,Seles2021}, are considered here
\begin{equation}
    \begin{split}
        f_0(\bar{\alpha})&={\left(1-\dfrac{\bar{\alpha}-\bar{\alpha}_{0}}{\bar{\alpha}+\bar{\alpha}_{0}}\right)^{2}} \hspace{7mm}\text{for}\hspace{2mm} \bar{\alpha} \in \left[\bar{\alpha}_{0},\,\infty\right]\;(\text{otherwise}\:f_0(\bar{\alpha})=1) \\[3mm]
        f_1(\bar{\alpha})&={\left(1-\dfrac{\bar{\alpha}}{\bar{\alpha}+\bar{\alpha}_{0}}\right)^{2}} \hspace{7mm}\text{for}\hspace{2mm} \bar{\alpha} \in \left[0,\,+\infty\right] \\[3mm]
        f_2(\bar{\alpha})&={\left(1-\dfrac{\bar{\alpha}}{\bar{\alpha}_{0}}\right)^{2}} \hspace{15mm}\text{for}\hspace{2mm} \bar{\alpha} \in \left[0,\,\bar{\alpha}_{0}\right] 
    \end{split}
    \label{eq:Fdegs}
\end{equation}
where $\bar{\alpha}_{0}$ is meant to be a material parameter to be calibrated with experiments. As shown in Fig. \ref{fig:Fdegs}, the main difference between them is that $f_0$ and $f_1$ deliver an asymptotically vanishing value while $f_2$ vanishes for a finite value of $\bar{\alpha}$. In addition, $f_0$ provides an initial threshold branch where material toughness remains unaffected by fatigue as the value of $\bar{\alpha}$ increases.

In addition, the fatigue history variable $\bar{\alpha}$ should describe the accumulation of any quantity $\alpha$ that can describe the cyclic history of the material. We follow Carrara et al. \cite{Carrara2020} in maintaining the energetic nature of the model and thus use the active part of the stored elastic energy density, defined in Section \ref{sec:StrainEnergyDecomposition}, as the fatigue history variable, i.e.
\begin{equation}
    \alpha=g(\phi)\psi^{+}_{0}(\bm{\varepsilon})
    \label{eq:alph}
\end{equation}

Note that the adoption of the \emph{degraded} strain energy density ensures that the quantity is not affected by the crack tip singularity. Accordingly, the evolution of the fatigue history variable $\bar{\alpha}$, within the time discretization, is given by
\begin{equation}
    \bar{\alpha}_{t+\Delta t}=\bar{\alpha}_{t}+\int_{t}^{{t+\Delta t}} \dot{\bar{\alpha}} \, \mathrm{d}\tau = \bar{\alpha}_{t} + \Delta \bar{\alpha}
    \label{eq:alphB}
\end{equation}

\begin{figure}[H]
    \centering
    \includegraphics[width=0.6\linewidth]{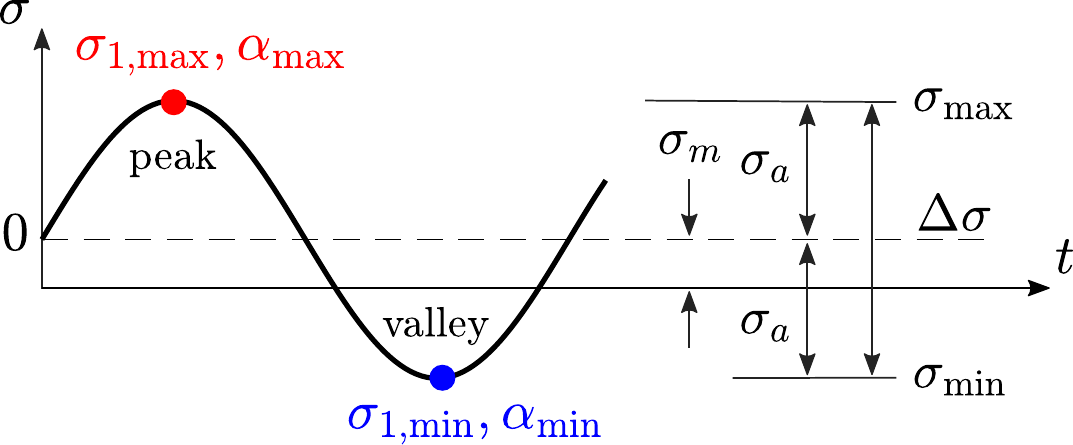}
    \caption{Constant amplitude cyclic stressing and definitions of the main variables. The red dot (peak) shows the location where $\sigma_{\mathrm{max}}^\mathrm{I}$ and $\alpha_{\text{max}}$ are calculated, where the blue dot (valley) shows the instant at which $\sigma_{\mathrm{min}}^\mathrm{I}$ and $\alpha_{\text{min}}$ are determined.}
    \label{fig:cyclic_stress}
\end{figure}

A key aspect in developing a constitutive phase field fatigue model lies in the definition of $\Delta \bar{\alpha}$; the approach employed to account for the accumulation of fatigue damage. In Ref. \cite{Carrara2020}, the accumulation of fatigue damage is considered only during the loading part of the cycle, which undesirably affects the proportional (monotonic) loading case. To address this issue, Seles et al. \cite{Seles2021} considered the accumulation of fatigue effects only during the unloading stage. However, we have observed that this might result in an unrealistic increase of the fatigue history variable in areas behind the crack tip as a result of localised unloading in those material points. Here, we suggest accumulating fatigue effects only during one reversal per cycle (peak to valley, see Fig. \ref{fig:cyclic_stress}), thus not affecting the monotonic loading cases. Most importantly, the new accumulation strategy enables us to achieve very significant reductions in computational cost as it allows us to accurately describe the accumulation of $\bar{\alpha}$ by using only one increment per cycle. Thus, for constant amplitude cases, internal increments within a cycle are instead replaced by the application of a constant (representative) load with the maximum value of the amplitude as its magnitude. As shown in Fig. \ref{fig:cyclic_stress}, the maximum and minimum values of the fatigue history variable are respectively denoted as $\alpha_{\text{max}}$ and $\alpha_{\text{min}}$, and can be estimated at the cycle peak and the valley during one reversal. 

Building upon our fatigue accumulation strategy, we proceed to define $\Delta \bar{\alpha}$ to present a model that accounts for (i) the slope of the S-N curve, (ii) the endurance limit, and (iii) the effect of the stress ratio. This generalised expression reads:
\begin{equation}
    \Delta \bar{\alpha}=\left(\frac{\alpha_{\text{max}}}{\alpha_n}\right)^n \left(\frac{1-R}{2}\right)^{2\kappa n} H\left(\max _{\tau \in[0, t]} \alpha_{\mathrm{max}} \left(\frac{1-R}{2}\right)^{2\kappa}-\alpha_e\right)
    \label{eq:dalphB_3}
\end{equation}
and each of its elements is described below. Here, one should note that $\Delta \bar{\alpha}$ is defined as a dimensionless quantity. A comparison with some of the main existing phase field fatigue models is provided in \ref{Sec:appendixA}.\\

\noindent \emph{S-N curve slope}. We add a material parameter, the exponent $n$, and an additional term, $(\alpha_{\text{max}}/\alpha_n)^n$, to endow the model with the flexibility needed to match the slope of any S-N curve. Here, a normalization parameter $\alpha_n$ is needed to achieve dimensional consistency. We adopt $\alpha_n=1/2\sigma_c\varepsilon_c$, based on the critical stresses and strains given in Eq. (\ref{eq:sigmac}).\\

\noindent \emph{Endurance limit}. A fatigue threshold variable $\alpha_e$ is introduced to endow the model with a material endurance limit, below which cyclic damage does not occur. This is used in combination with the Heaviside function $H\left(\square\right)$, which equals one for positive arguments and zero for negative ones. The magnitude of $\alpha_e$ can be estimated from the material endurance stress $\sigma_e$ as $\alpha_e=\sigma_e^2/(2E)$.\\

\noindent \emph{Stress ratio}. Fatigue behaviour is known to exhibit significant sensitivity to the stress ratio, which can be defined as $R=\sigma_{1,\mathrm{min}} / \sigma_{1,\mathrm{max}}$, where $\sigma_{1,\mathrm{min}}$ and $\sigma_{1,\mathrm{max}}$ respectively denote the minimum and maximum principal stresses within each cycle (see Fig. \ref{fig:cyclic_stress}). In the case of proportional loading, which is the case for all analyses in the present paper, this does not lead to ambiguity. However, for non-proportional loading care must be taken to define the values properly. A suitable choice may be to choose the direction, $\bm{n}_1$, according to the maximum principle value, and evaluate both the maximum and the minimum normal stresses in this direction. It should be noted that $R$ is not an input to the model but a material point quantity that can be estimated at the end of each cycle. To introduce $R$ into the accumulation of the fatigue history variable, we take inspiration from classical mean stress relationships. In particular, the Walker mean stress relationship \cite{Walker1970} has been widely used to enrich Basquin-type laws to account for non-zero mean stresses; this relationship reads,
\begin{equation}
    \sigma_{ar}=\sigma_{\mathrm{max}}\left(\frac{1-R}{2}\right)^{\kappa},  \hspace{7mm} \text{for} \hspace{2mm} (\sigma_{\mathrm{max}}>0)
\end{equation}
where $\sigma_{ar}$ is the equivalent stress amplitude when the mean stress is $\sigma_m=0$, $\sigma_{\mathrm{max}}$ is the maximum stress within each cycle, and $\kappa \in [0, 1]$ is a material constant, describing the measure of the material’s sensitivity to mean stress. For $\kappa=0.5$, the Walker equation reduces to the well-known Smith-Watson-Topper (SWT) relationship \cite{Smith1970}. As shown in Eq. (\ref{eq:dalphB_3}), our model employs Walker-based terms to capture the load ratio effect. Other approaches, involving the use of sign functions (see \ref{Sec:appendixA}), did not provide a good agreement with experiments.

\section{Numerical implementation}
\label{Sec:Numerical}
Details of the numerical implementation are provided here, starting with the finite element discretization
(Section \ref{Sec:Discretization}), followed by the formulation of the residuals and the stiffness matrices (Section \ref{Sec:RHS_K}).

\subsection{Finite element discretization}
\label{Sec:Discretization}
The finite element (FE) method is used to solve the coupled problem. Making use of Voigt notation, the primary kinematic variables of the coupled problem are discretized in terms of their nodal values $\textbf{u}_{i}=\left\{u_x, u_y, u_z\right\}^{\mathsf{T}}_{i}$ and $\phi_{i}$ at node $i$ as

\begin{equation}
    \mathbf{u}=\sum_{i=1}^m \mathbf{N}_i \mathbf{u}_i \hspace{10mm}\text{and}\hspace{10mm}  \phi=\sum_{i=1}^m N_i \phi_i
    \label{eq:Discret}
\end{equation}
where $m$ is the total number of nodes per element, $N_{i}$ the shape functions associated with node $i$, and $\mathbf{N}_i$ the shape function matrix, a diagonal matrix with $N_i$ in the diagonal terms. Accordingly, the corresponding gradient quantities can be discretized as
\begin{equation}
    \bm{\varepsilon}=\sum_{i=1}^m \mathbf{B}^u_i \mathbf{u}_i \hspace{10mm}\text{and}\hspace{10mm}   \nabla \phi=\sum_{i=1}^m \mathbf{B}_i \phi_i
    \label{eq:GradDiscret}
\end{equation}
where $\mathbf{B}_{i}^{\mathbf{u}}$ denotes the standard strain-displacement matrices and  $\mathbf{B}_{i}$ is a vector containing the spatial derivatives of the shape functions.

\subsection{Residuals and stiffness matrices}
\label{Sec:RHS_K}
We now proceed to formulate the weak form of the coupled problem. Considering the principle of virtual power (\ref{eq:VWP}) and the constitutive choices described in Section \ref{sec:ConstitutiveTheory}, the weak forms of the displacement and phase field problems read
\begin{equation}
\begin{split}
    \int_{\Omega}& \bigg\{ \big[g(\phi)+k\big] \bm{\sigma}_0: \nabla \dot{\mathbf{u}} - \mathbf{b} \cdot \dot{\mathbf{u}} \bigg\} \, \mathrm{d}V - \int_{\partial \Omega_{h}} \mathbf{h} \cdot \dot{\mathbf{u}}\, \mathrm{d}S = 0\\[3mm]
    \int_{\Omega}& \bigg\{ g^{\prime}(\phi)\dot{\phi} \, \mathcal{H} + f(\bar{\alpha})\frac{G_{c}}{4 c_{w}}\left(\frac{w^{\prime}(\phi)\dot{\phi}}{\ell}+2\ell\nabla \phi\cdot\nabla\dot{\phi}\right) \bigg\}  \, \mathrm{d}V - \int_{\partial \Omega_{f}} f\, \dot{\phi}\, \mathrm{d}S= 0
\end{split}
\label{eq:Weak}
\end{equation}
where $\bm{\sigma}_0$ is the Cauchy stress tensor of the undamaged solid and $k$ is a small and positive constant used to avoid ill-conditioning of the system of equations when $\phi=1$; in this work $k=10^{-7}$. Now, making use of the finite element discretization outlined in (\ref{eq:Discret}) and (\ref{eq:GradDiscret}) and considering that (\ref{eq:Weak}) must hold for any kinematically admissible variations of the virtual quantities $\dot{\square}$, the corresponding residuals are derived as
\begin{equation}
\begin{split}
    \mathbf{r}_{i}^u=&\int_{\Omega} \big[g(\phi)+k\big] {(\mathbf{B}_{i}^{u})}^{\mathsf{T}} \bm{\sigma_{0}} \, \mathrm{d}V - \int_{\Omega} {(\mathbf{N}_{i})}^{\mathsf{T}} \mathbf{b} \, \mathrm{d}V - \int_{\partial \Omega_{h}} {(\mathbf{N}_{i})}^{\mathsf{T}} \mathbf{h} \, \mathrm{d}S\\[3mm]
    r_{i}^{\phi}=& \int_{\Omega} \left\{ g^{\prime}(\phi) N_{i} \, \mathcal{H} +f(\bar{\alpha})\frac{G_{c}}{4 c_{w}}\left(\frac{w^{\prime}(\phi)}{\ell} N_{i}+ 2\ell {(\mathbf{B}_{i})}^{\mathsf{T}} \nabla \phi \right) \right\} \, \mathrm{d}V - \int_{\partial \Omega_{f}} N_{i} \, f\, \mathrm{d}S
\end{split}
\label{eq:{RHS}}
\end{equation}

Finally, the consistent tangent stiffness matrices are obtained by differentiating the residuals with respect to the incremental nodal variables as follows
\begin{equation}
\begin{split}
    \mathbf{K}_{ij}^{u} &= \dfrac{\partial \mathbf{r}_{i}^u}{\partial \mathbf{u}_{j}} = \int_{\Omega} \big[g(\phi)+k\big] {(\mathbf{B}_{i}^{u})}^{\mathsf{T}} \bm{\mathcal{L}}_{0}  \, \mathbf{B}_{j}^{u} \, \mathrm{d}V \\[3mm]
    \mathbf{K}_{ij}^{\phi} &=  \dfrac{\partial r_{i}^{\phi}}{\partial \phi_{j}} = \int_{\Omega} \left\{ \left( g^{\prime\prime}(\phi)\,\mathcal{H} + f(\bar{\alpha})\dfrac{G_{c}}{4c_w\ell}w^{\prime\prime}(\phi) \right) N_{i} N_{j} + f(\bar{\alpha})\dfrac{G_{c}\ell}{2c_w} {(\mathbf{B}_{i})}^{\mathsf{T}} \mathbf{B}_{j} \right\} \, \mathrm{d}V
\end{split}
\label{eq:[K]}
\end{equation}

We then solve the global linearized FE system of equations,
\begin{equation}
    \begin{bmatrix}
    {\mathbf{K}^{u}} & {0} \\[0.3em] 
    {0} & {\mathbf{K}^{\phi}}
    \end{bmatrix}
    \begin{Bmatrix}
    {\mathbf{u}}\\[0.3em] 
    {\bm{\phi}}
    \end{Bmatrix}
    =\begin{Bmatrix}
    {\mathbf{r}^{u}}\\[0.3em]
    {\mathbf{r}^{\phi}}
    \end{Bmatrix}
    \label{eq:FEsystemEqs}
\end{equation}
by using a quasi-Newton method. Specifically, we employ the Broyden–Fletcher–Goldfarb–Shanno (BFGS) algorithm \cite{Wu2020c,Kristensen2020}, which provides a robust monolithic solution scheme, enabling accurate and efficient fatigue crack growth estimations. Note that, a requirement of the BFGS algorithm is that the stiffness matrix must be symmetric and positive-definite.

\section{Results}
\label{Sec:Results}

\subsection{Smooth bar subjected to symmetric uniaxial tension-compression loading}
We first gain insight into the model characteristics by considering a smooth bar subjected to uniaxial cyclic loading with a load ratio of $R=-1$. A model material is assumed with the following properties: Young’s modulus $E=1$ MPa, Poisson’s ratio $\nu=0.3$, tensile strength $\sigma_c=1$ MPa, endurance limit $\sigma_e=0.2$ MPa, critical energy release rate $G_c=1$ kJ/m$^2$ and fatigue material parameter $\bar{\alpha}_0=100$. The boundary value problem can be solved in a semi-analytical fashion, by considering the homogeneous solution to Eq. (\ref{eq:StrongFormPhi}). A piece-wise cyclic linear variation of the remote stress (or strain) is assumed. Under 1D conditions, the length scale and the strength are related via (\ref{eq:sigmac}), and this relation renders magnitudes of $\ell=0.3750$ mm and $\ell=0.1055$ mm for \texttt{AT1} and \texttt{AT2}, respectively. Unless otherwise stated, in the remainder of this paper the \texttt{AT1} model, $\kappa=0.5$, the $f_2$ fatigue degradation function (\ref{eq:Fdegs}c) and the \texttt{No-tension} split (\ref{eq:decFreddi}) are used. While all the numerical studies conducted deal with constant amplitude loading, we emphasise that the model can handle any arbitrary choice of loading history and thus capture load sequence effects.

\subsubsection{Overview of material behaviour}

\begin{figure}[hbt!]
\begin{subfigure}[t]{0.32\textwidth}
        \centering
    \includegraphics[width=\textwidth,trim={0 0 0 0},clip]{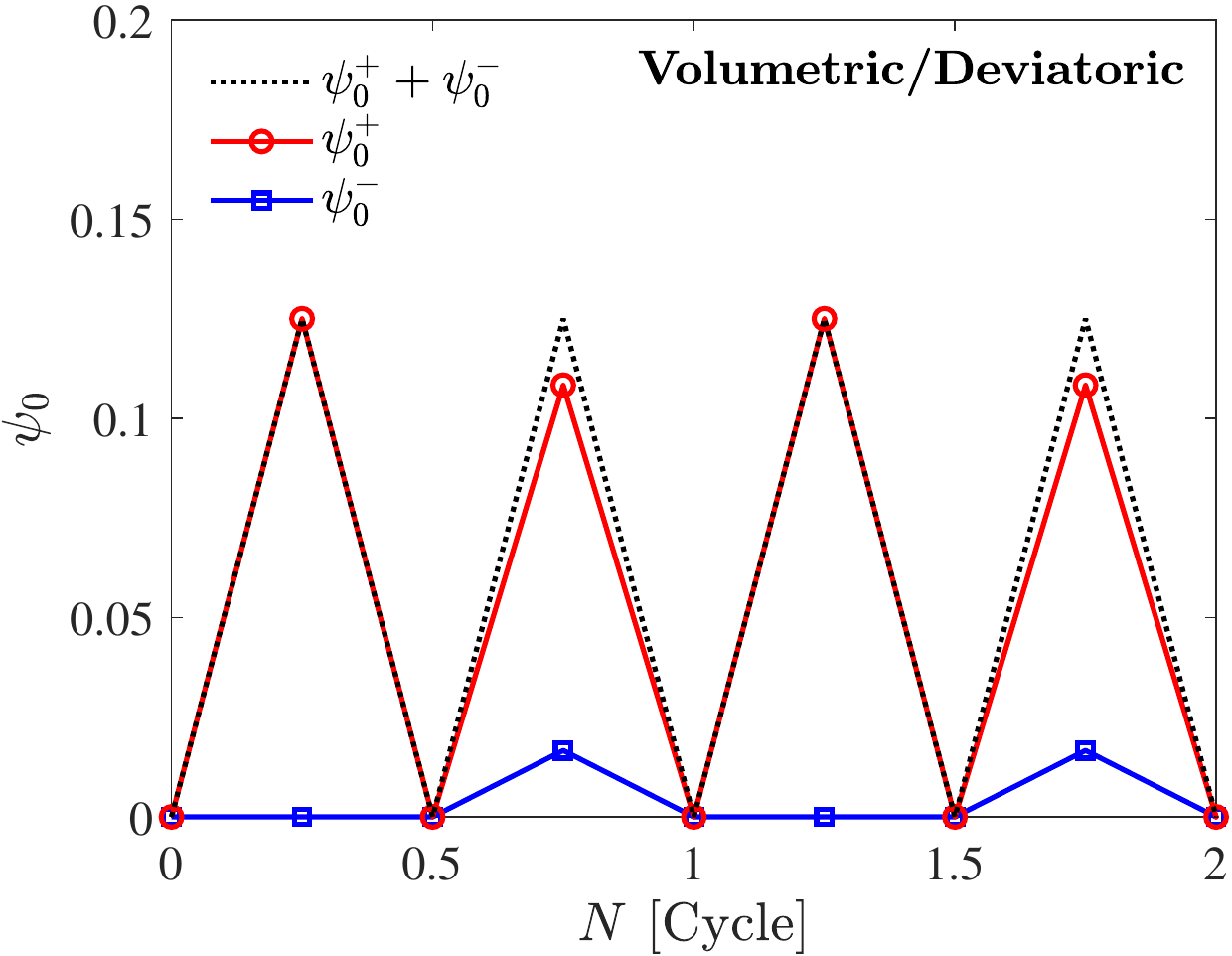}
    \caption{}
    \end{subfigure}
    \begin{subfigure}[t]{0.32\textwidth}
        \centering
    \includegraphics[width=1\linewidth,trim={0 0 0 0},clip]{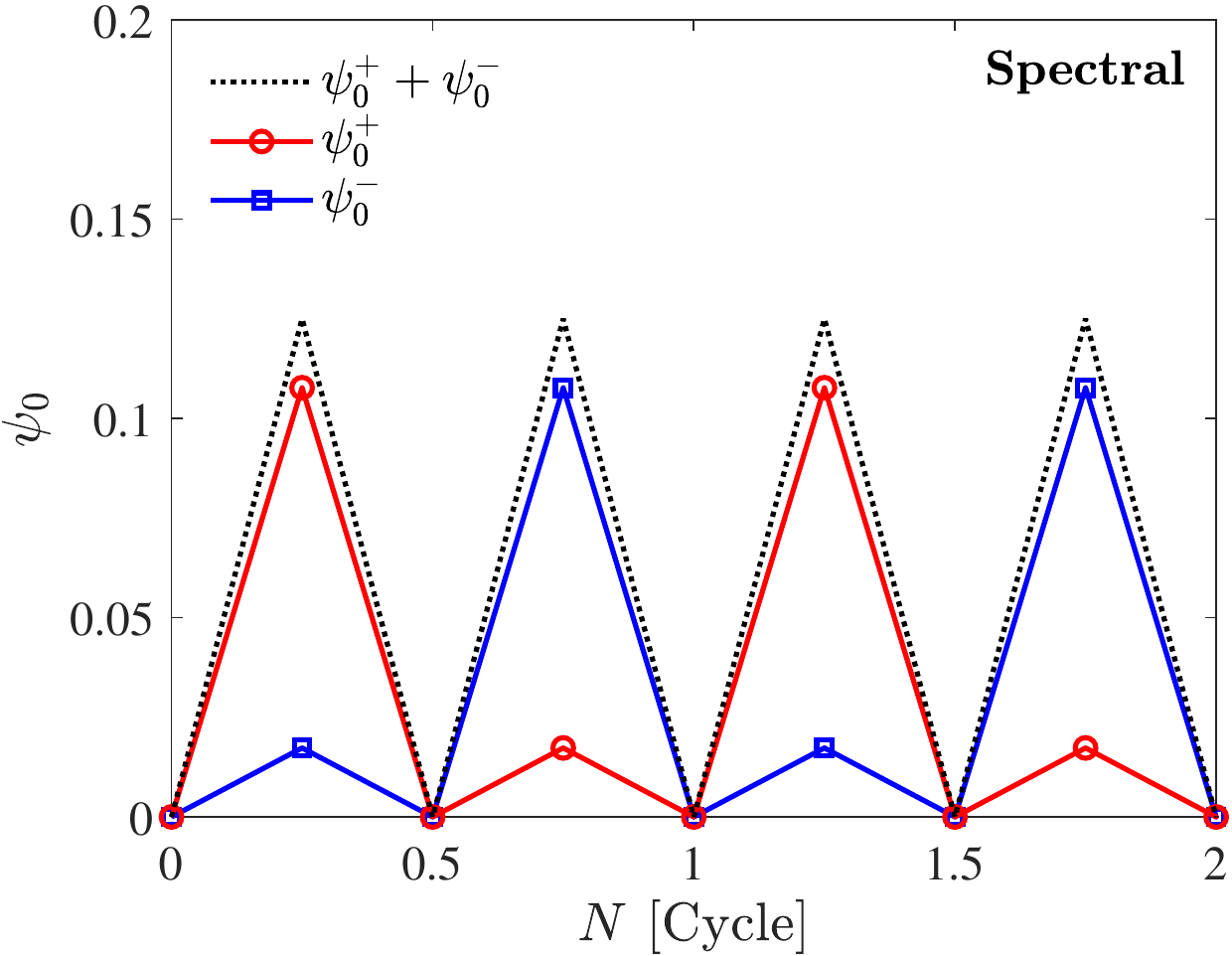}
    \caption{}
    \end{subfigure}
\begin{subfigure}[t]{0.32\textwidth}
        \centering
    \includegraphics[width=1\linewidth,trim={0 0 0 0},clip]{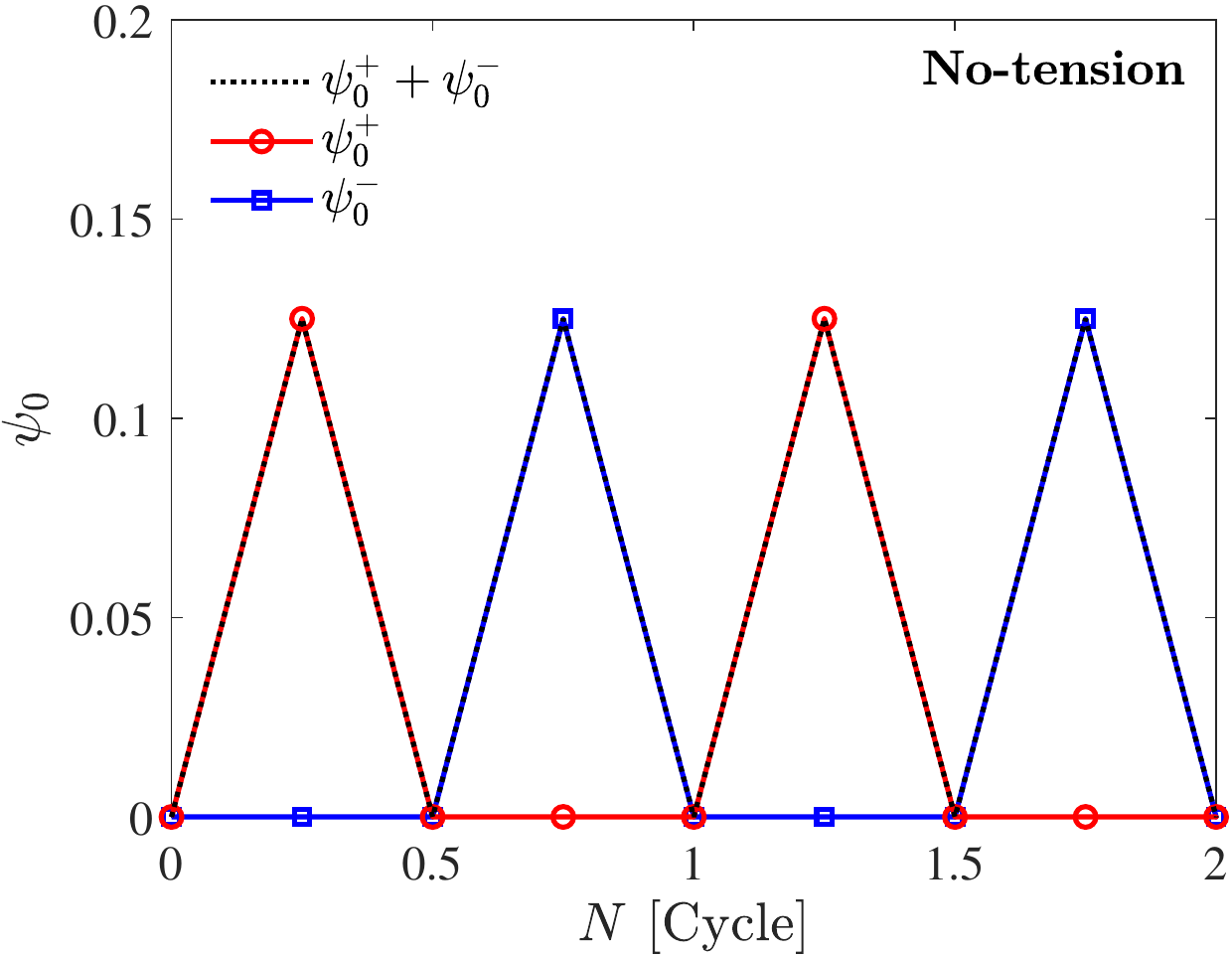}
    \caption{}
    \end{subfigure}
    \caption{Sensitivity of fatigue driving force, recall $\alpha=g(\phi)\psi^{+}_{0}$, to the choice of strain energy density decomposition; tensile $\psi_0^+$ and compressive $\psi_0^-$ components for a fully-reversed cyclic loading ($R=-1$) considering the (a) \texttt{Volumetric/deviatoric}, (b) \texttt{Spectral}, and (c) \texttt{No-tension} splits.}
    \label{fig:DecompositionSplit}
\end{figure}
\begin{figure}[ht!]
\begin{subfigure}[t]{0.49\textwidth}
        \centering
    \includegraphics[width=0.96\linewidth,trim={0 0 0 0},clip]{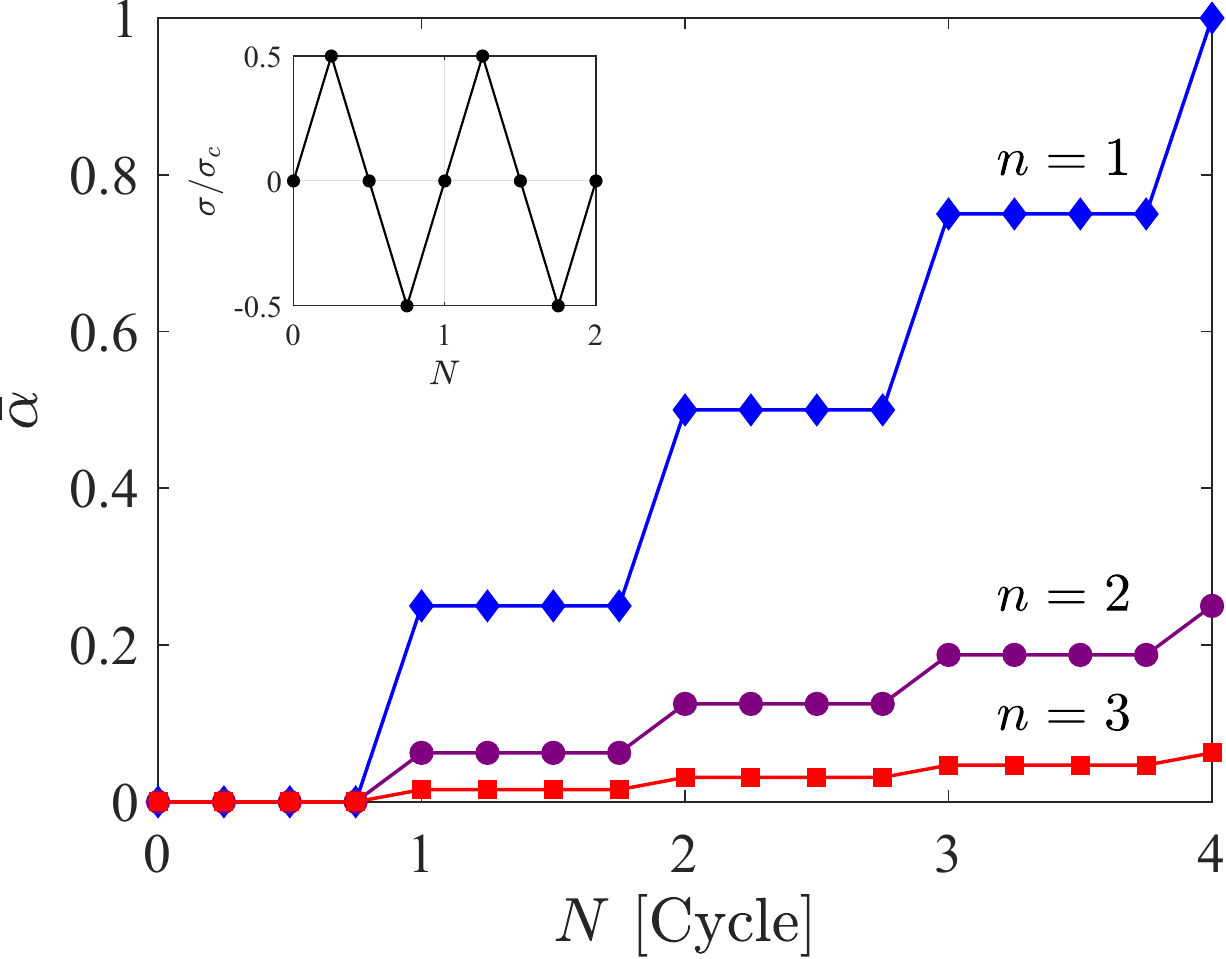}
    \caption{}
    \end{subfigure}
        \begin{subfigure}[t]{0.49\textwidth}
        \centering
    \includegraphics[width=1\linewidth,trim={0 0 0 0},clip]{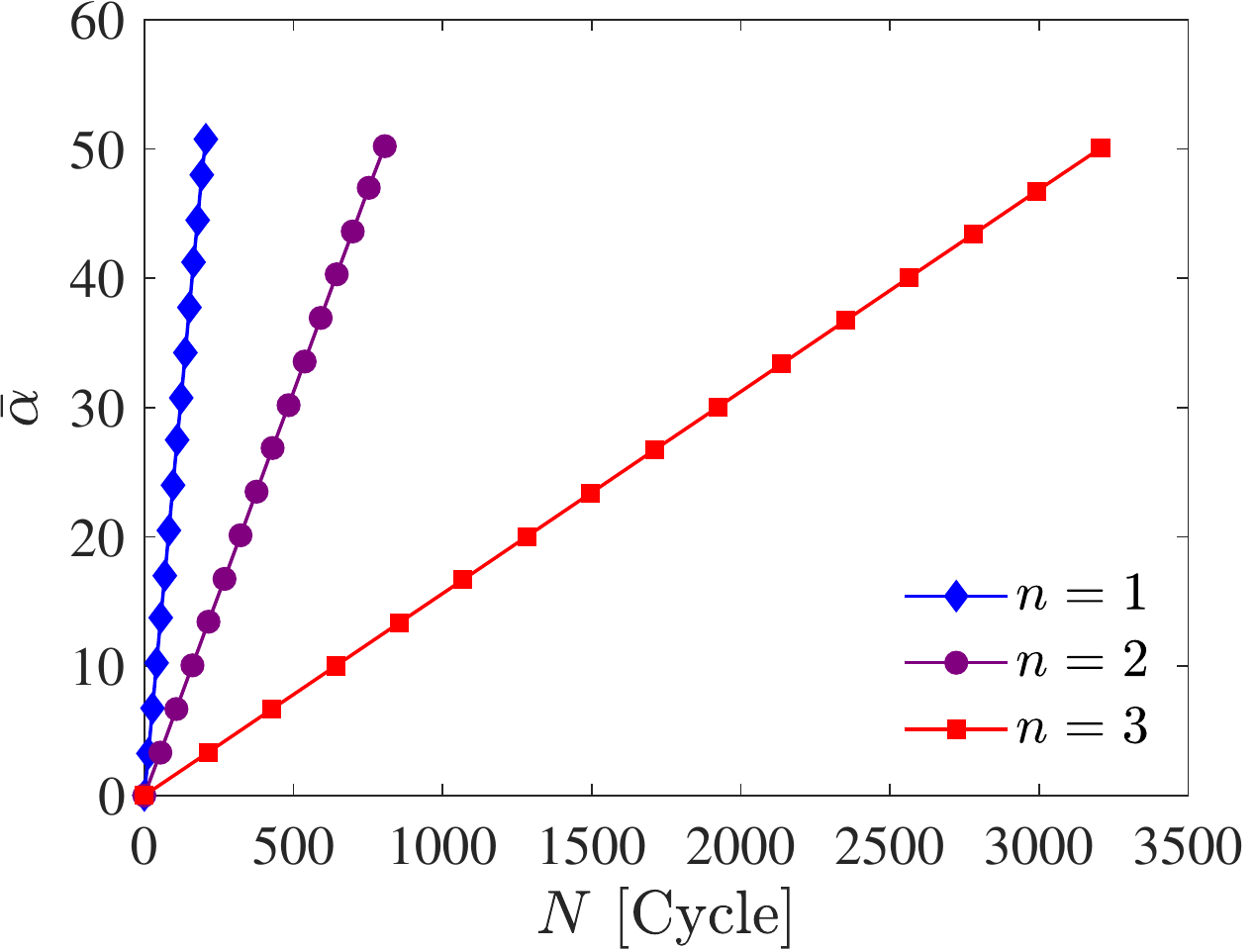}
    \caption{}
    \end{subfigure}
    \caption{Cyclic evolution of the fatigue history variable $\bar{\alpha}$ for different values of the power exponent $n$: (a) detail of the first cycles, showing how the \texttt{No-tension} split appropriately accumulates damage only within one half-cycle per cycle, and (b) evolution over numerous cycles, showing the influence of the exponential parameter $n$.}
    \label{fig:AlphB_vs_N}
\end{figure}

Fig. \ref{fig:DecompositionSplit} illustrates the evolution of the elastic strain energy density along with its active (tensile) and inactive (compressive) parts for a constant remote stress amplitude of $\sigma_a/\sigma_c=0.5$, upon the assumption of a fatigue power exponent of $n=1$. It can be clearly seen that the \texttt{No-tension} split appropriately decomposes the strain energy density such that it results in a vanishing compressive part during tension and a vanishing tensile part during compression, which is not the case for the \texttt{Volumetric/deviatoric} and the \texttt{Spectral} splits. The consistency of the \texttt{No-tension} split is also showcased in Fig. \ref{fig:AlphB_vs_N}, where the cyclic evolution of the fatigue history variable $\bar{\alpha}$ is shown. It can be seen that the accumulation of fatigue effects takes place only during the reversal (peak to valley) part of each cycle, and that the growth rate of $\bar{\alpha}$ decreases with increasing the power exponent $n$.

\begin{figure}[ht!]
\begin{subfigure}[t]{0.49\textwidth}
        \centering
    \includegraphics[width=0.92\linewidth,trim={0 0 0 0},clip]{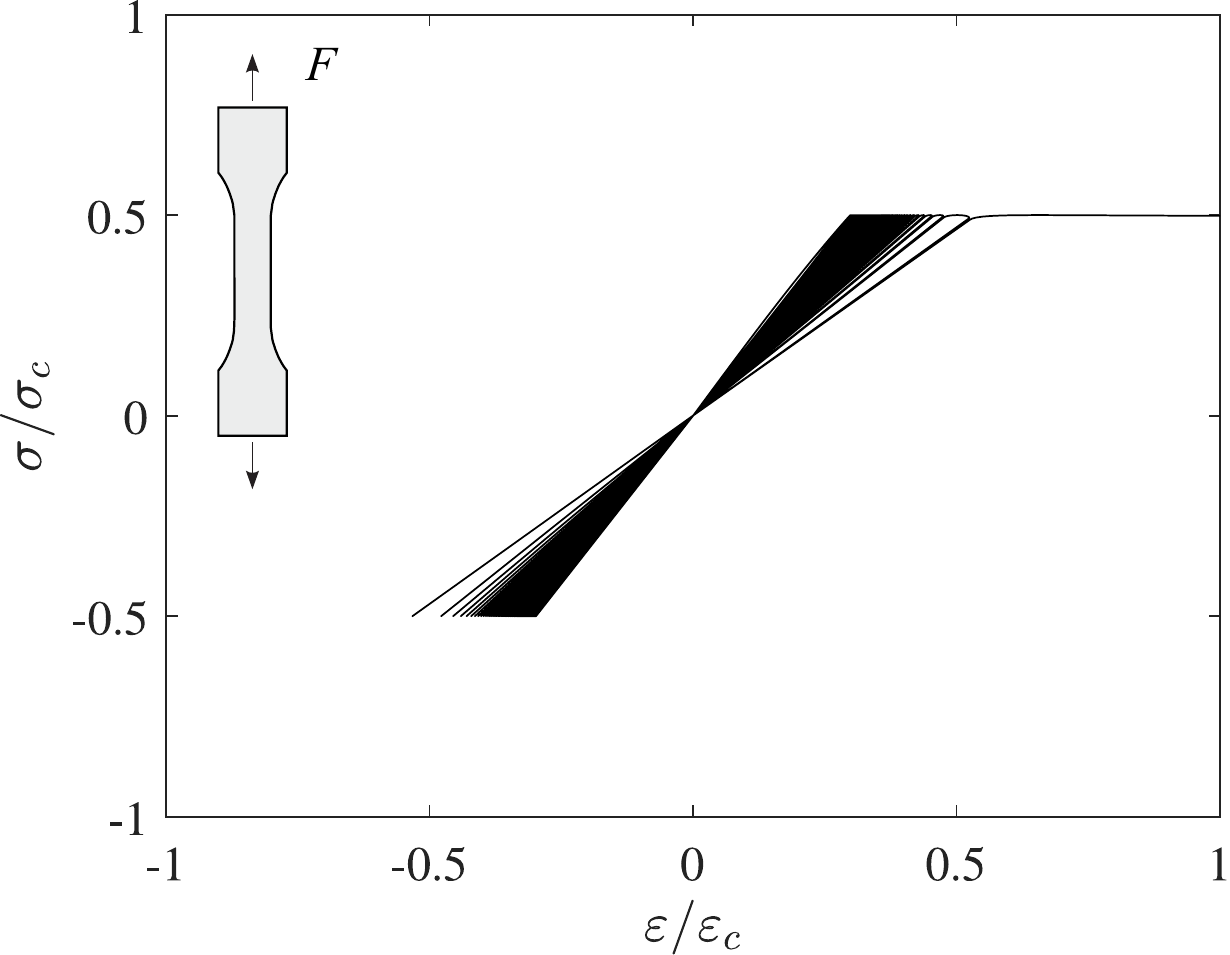}
    \caption{}
    \end{subfigure}
        \begin{subfigure}[t]{0.49\textwidth}
        \centering
    \includegraphics[width=1\linewidth,trim={0 0 0 0},clip]{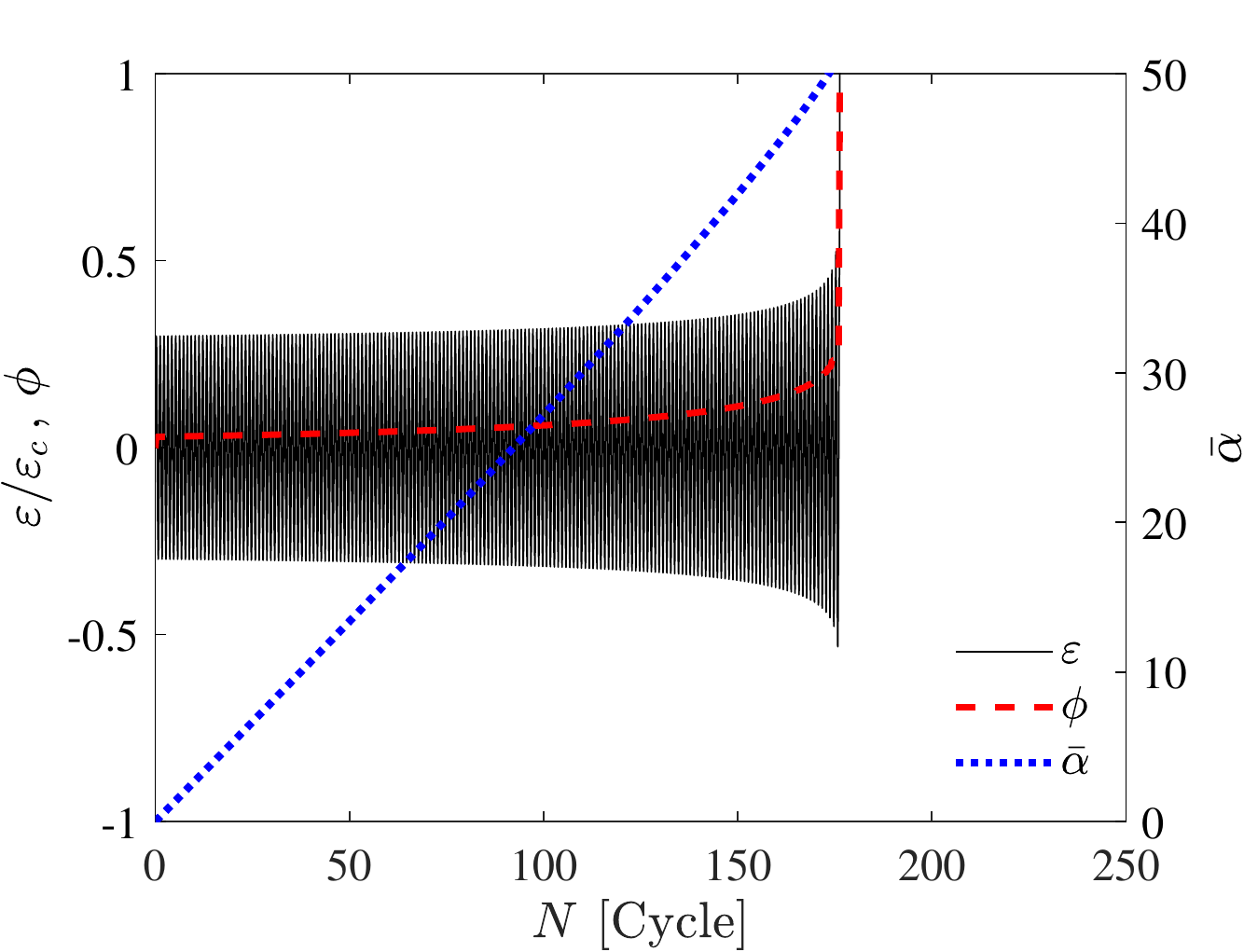}
    \caption{}
    \end{subfigure}
\\[3mm]
\begin{subfigure}[t]{0.49\textwidth}
        \centering
    \includegraphics[width=0.92\linewidth,trim={0 0 0 0},clip]{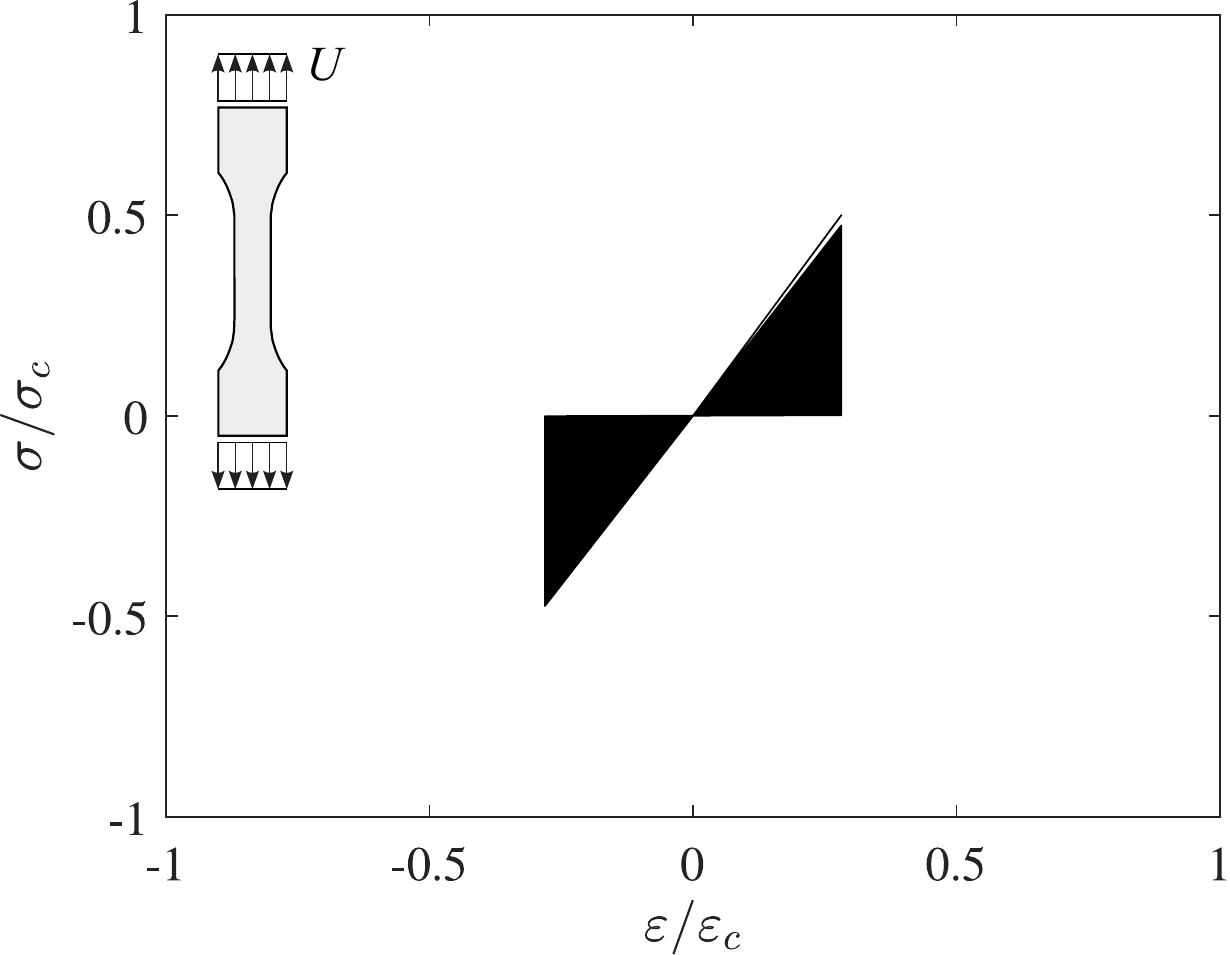}
    \caption{}
    \end{subfigure}
        \begin{subfigure}[t]{0.49\textwidth}
        \centering
    \includegraphics[width=1\linewidth,trim={0 0 0 0},clip]{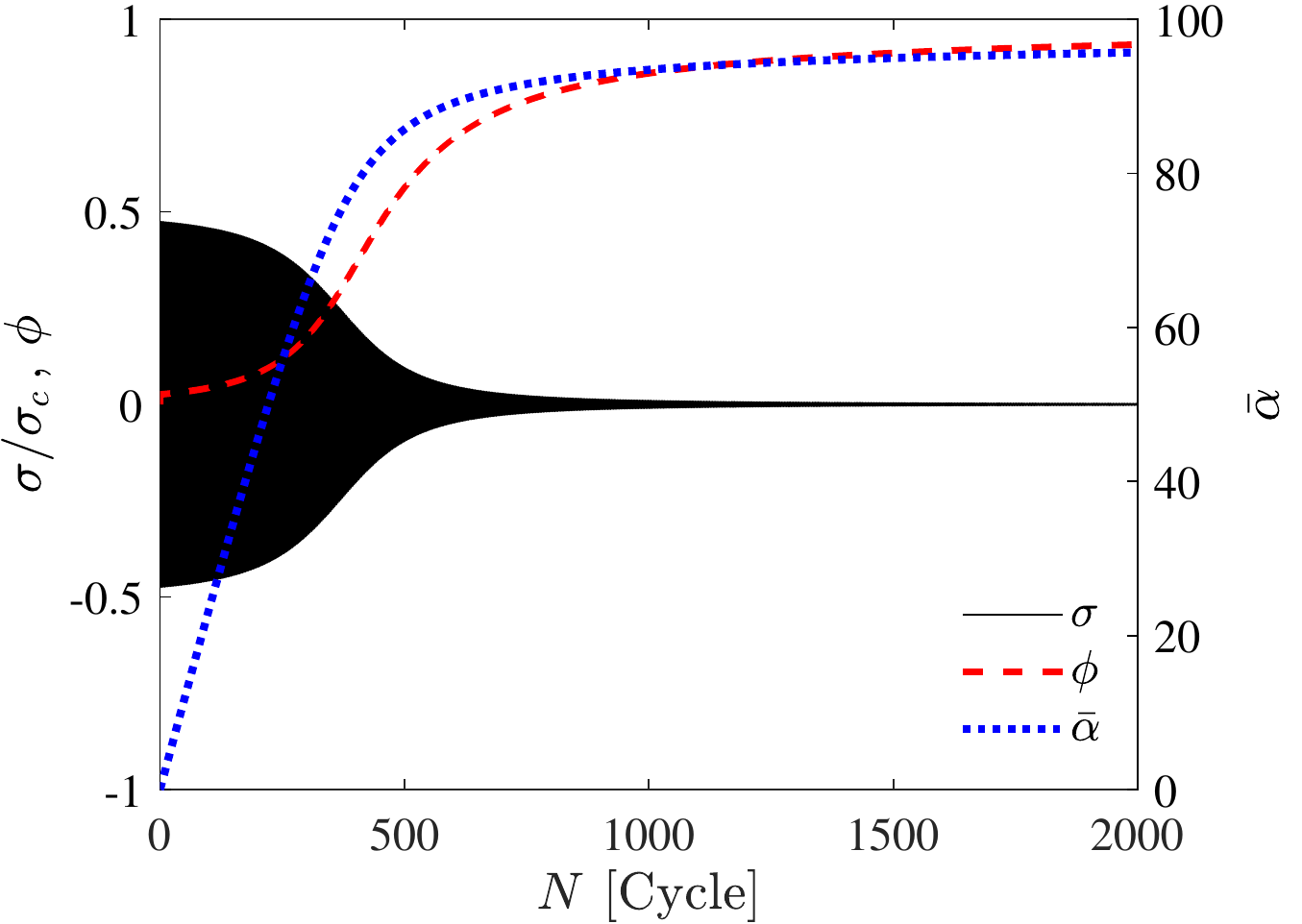}
    \caption{}
    \end{subfigure}
    \caption{Uniaxial tension-compression response under load-controlled, (a) and (b), and displacement-controlled conditions, (c) and (d). Stress versus strain curves are shown in (a) and (c), while (b) and (d) show the evolution of relevant variables ($\bar{\alpha}, \phi$, cyclic stress/strain) as a function of the number of cycles $N$. The number of cycles considered results in nearly overlapping curves (black regions). Calculations obtained using \texttt{AT2}, $n=1$, $\kappa=0.5$ and the \texttt{No-tension} split.}
    \label{fig:AT2_response}
\end{figure}
Further insight into the evolution of the model behaviour can be gained by comparing the differences between load-controlled and displacement-controlled numerical experiments. To this end, we use the \texttt{AT2} phase field model and conduct simulations: (i) applying a remote stress amplitude of $\sigma_a/\sigma_c=0.5$ (load-control), and (ii) applying a remote strain amplitude of $\varepsilon_a/\varepsilon_c=0.5$ (displacement-control). The results obtained are given in Figs. \ref{fig:AT2_response}(a)-(b) for load-controlled loading and in \ref{fig:AT2_response}(c)-(d) for displacement-controlled loading. These figures illustrate both material stress-strain behaviour and the evolution with the number of cycles ($N$) of relevant variables ($\bar{\alpha}, \phi$, cyclic stress/strain). As shown in Fig. \ref{fig:AT2_response}b, for the load-controlled case the phase field evolves gradually in the beginning and increases rapidly towards the end, when the strain reaches its critical value at $\varepsilon_c$. However, this is not the case for the displacement-controlled loading where the phase field is observed to asymptotically approach its upper limit $\phi\rightarrow1$ (see Fig. \ref{fig:AT2_response}d). Accordingly, a threshold for failure (e.g., $\phi=0.95$) must be imposed when considering displacement-control conditions. This variation of $\phi$ in time affects the cyclic evolution of the fatigue history variable $\bar{\alpha}$ as well as the cyclic stress, owing to the phase field degradation function (\ref{eq:gphi}), which is present in the definitions of $\bm{\sigma}$ (\ref{eq:CauchyStress}) and $\alpha$ (\ref{eq:alph}).

\begin{figure}[hbt!]
\begin{subfigure}[t]{0.49\textwidth}
        \centering
    \includegraphics[width=1\linewidth,trim={0 0 0 0},clip]{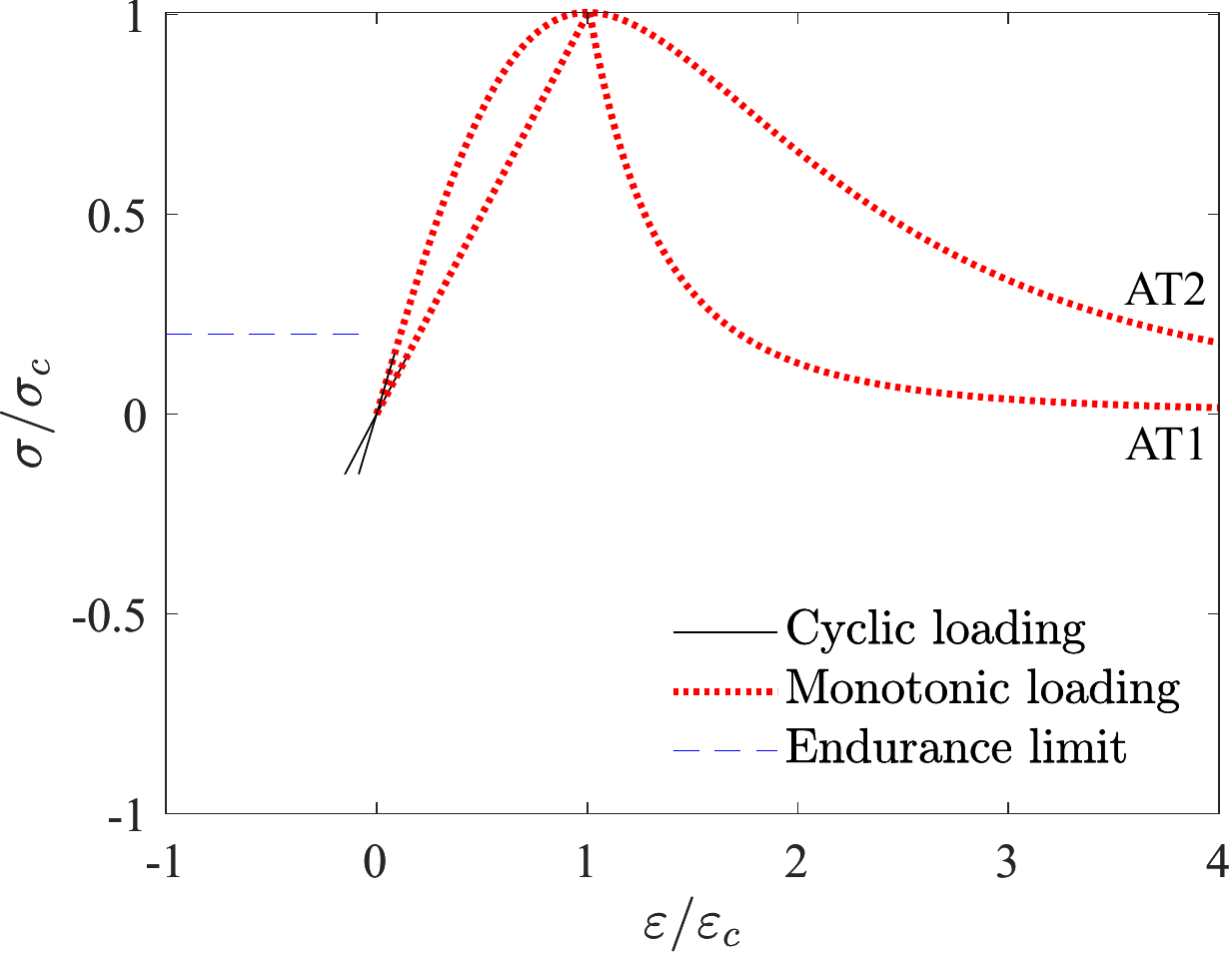}
    \caption{}
    \end{subfigure}
        \begin{subfigure}[t]{0.49\textwidth}
        \centering
    \includegraphics[width=1\linewidth,trim={0 0 0 0},clip]{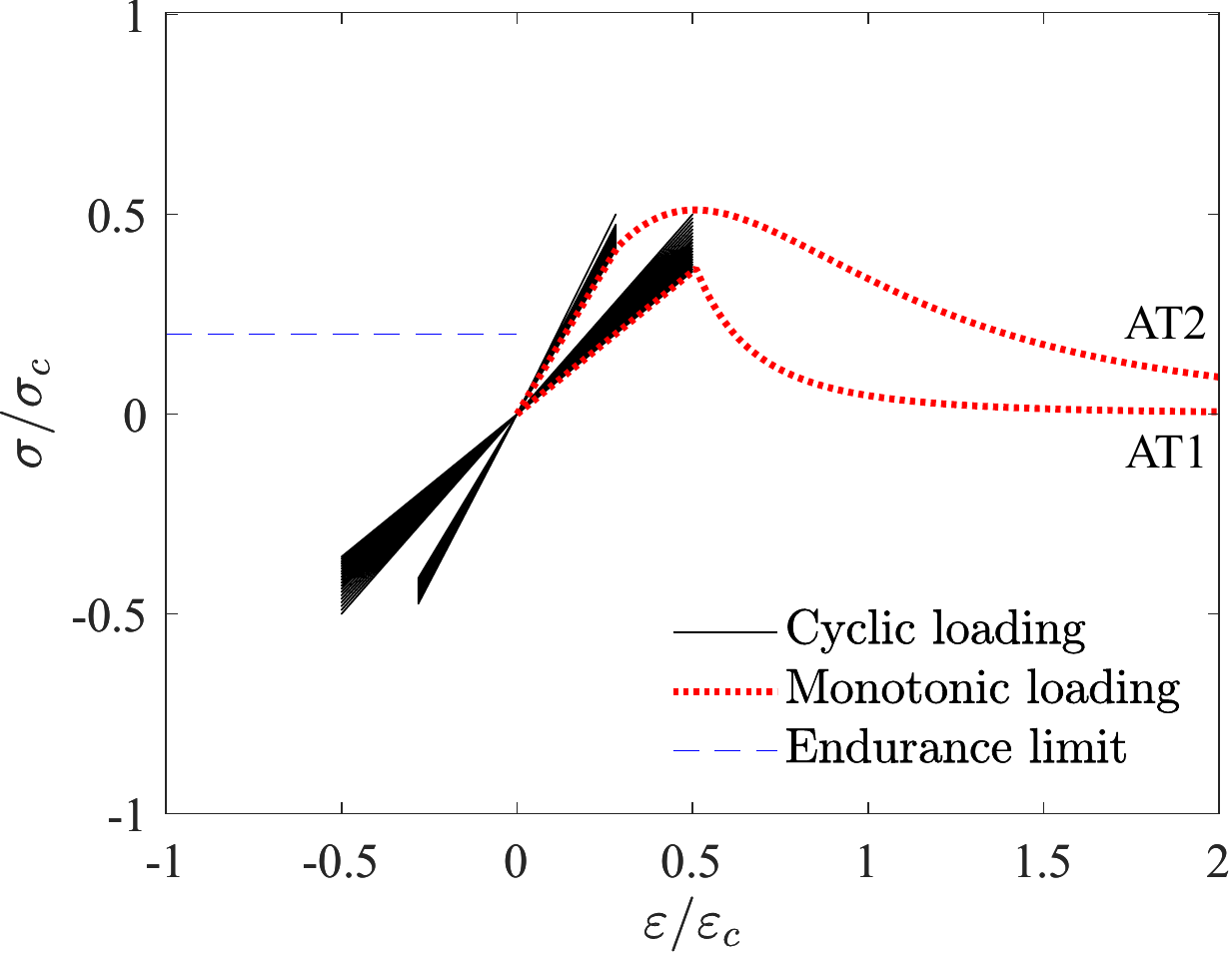}
    \caption{}
    \end{subfigure}
    \caption{Uniaxial cyclic and monotonic response of the \texttt{AT1} and \texttt{AT2} damage models, for different initially-applied remote strain amplitudes: (a) $\varepsilon_a/\varepsilon_c=0.15$ and (b) $\varepsilon_a/\varepsilon_c=0.5$. The number of cycles considered results in nearly overlapping curves (black regions). Calculations obtained using $n=1$, $\kappa=0.5$ and the \texttt{No-tension} split.}
    \label{fig:Mono_Cycle}
\end{figure}

We proceed to gain further insight by investigating the role of the phase field fracture constitutive model (\texttt{AT1} vs \texttt{AT2}) and the load amplitude ($\varepsilon_a/\varepsilon_c=0.15$ vs $\varepsilon_a/\varepsilon_c=0.5$). The results obtained are shown in Fig. \ref{fig:Mono_Cycle}. For the strain amplitude $\varepsilon_a/\varepsilon_c=0.15$, the resulting stresses are below the assumed material endurance limit ($\sigma_e/\sigma_c=0.2$) and as a result the monotonic response of the bar and its critical strength (strain) are not affected by fatigue (see Fig. \ref{fig:Mono_Cycle}a). On the other hand, when the load amplitude exceeds the endurance limit ($\varepsilon_a/\varepsilon_c=0.5$, Fig. \ref{fig:Mono_Cycle}b), the monotonic response of the bar exhibits a significant drop in the critical strength and strain of the bar. This is observed for both \texttt{AT1} and \texttt{AT2} models, being more significant in the former.

\subsubsection{Parametric study}

\begin{figure}[hbt!]
\begin{subfigure}[t]{0.49\textwidth}
        \centering
    \includegraphics[width=1\linewidth,trim={0 0 0 0},clip]{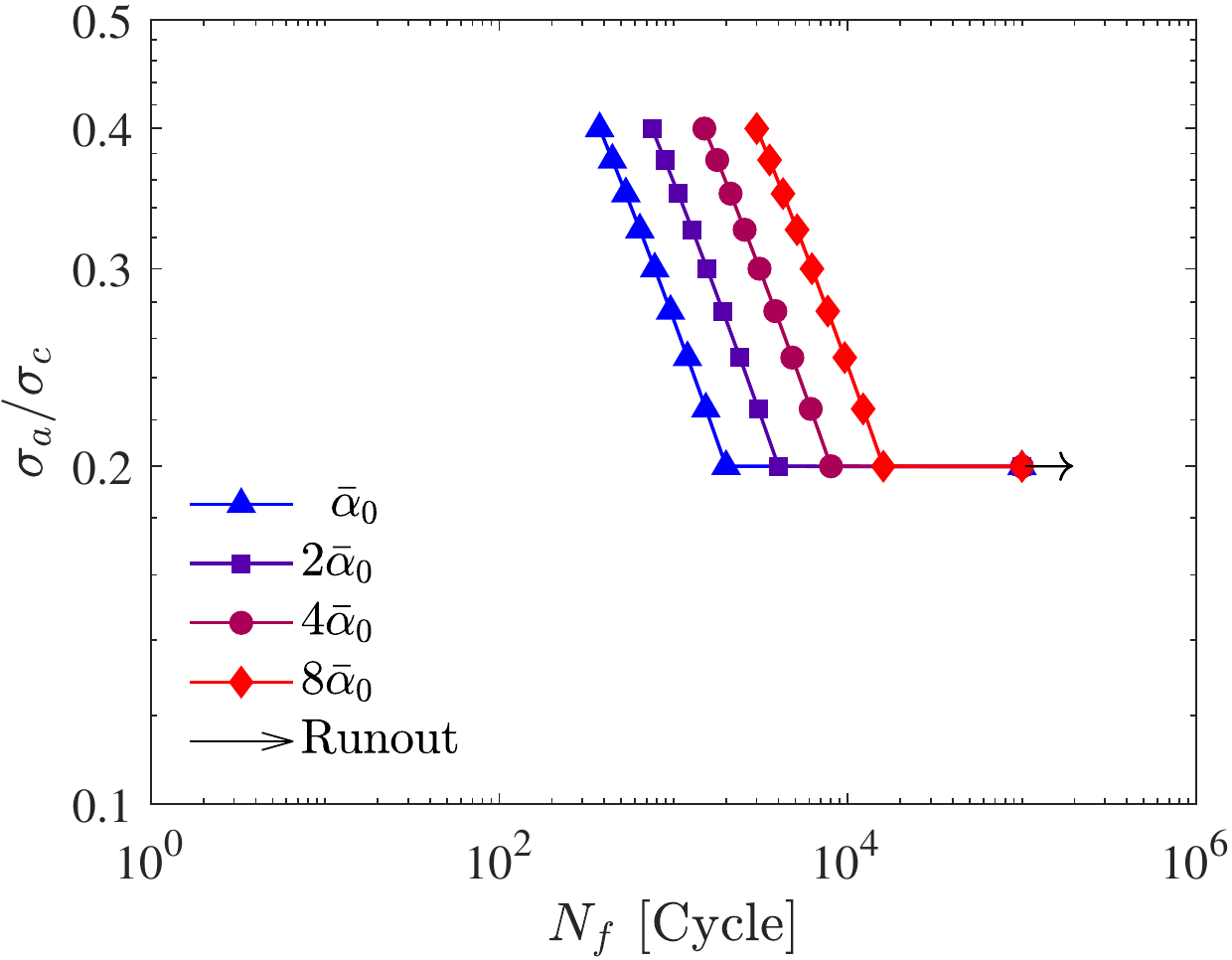}
    \caption{}
    \end{subfigure}
        \begin{subfigure}[t]{0.49\textwidth}
        \centering
    \includegraphics[width=1\linewidth,trim={0 0 0 0},clip]{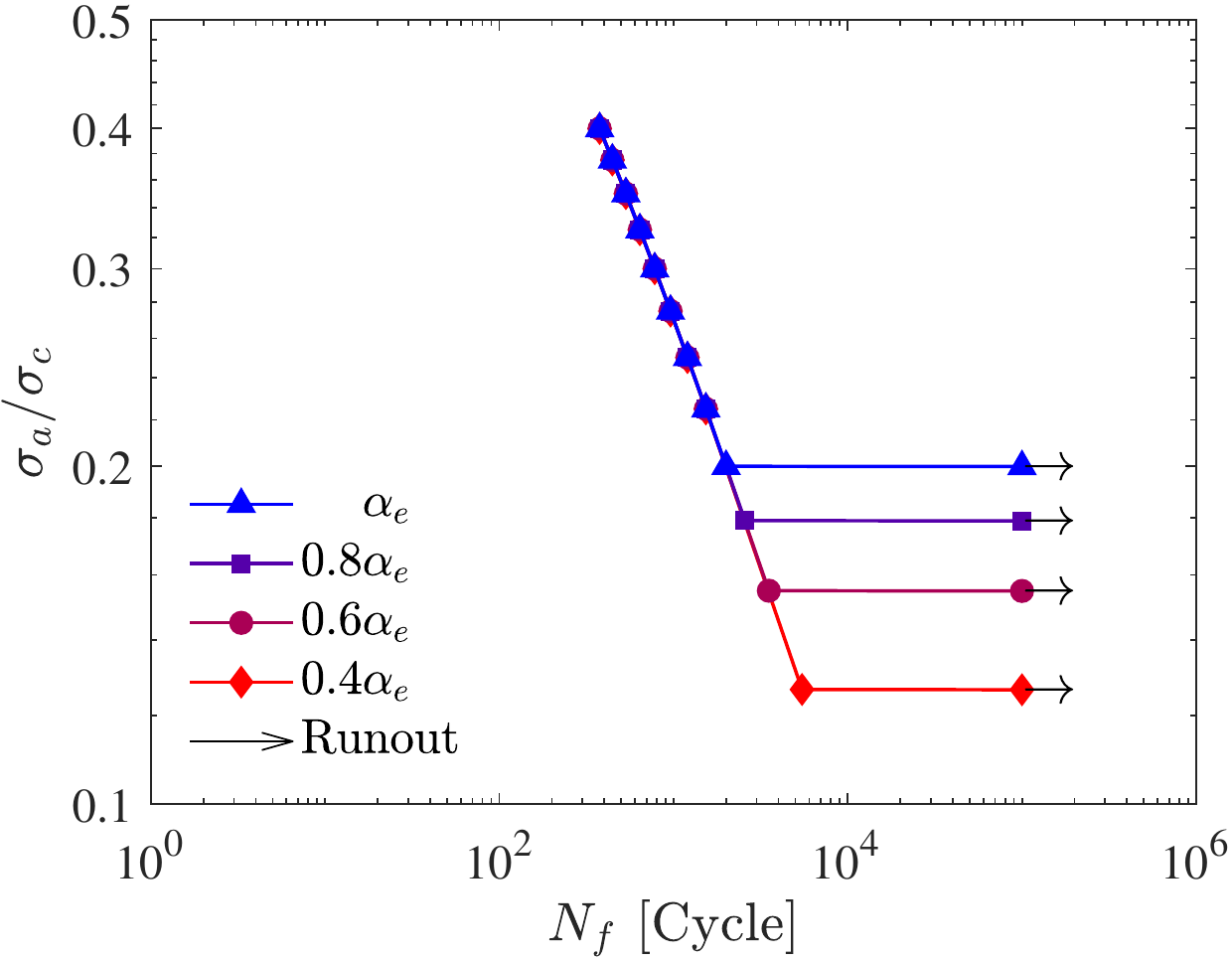}
    \caption{}
    \end{subfigure}
    \caption{Parametric study. S-N curve sensitivity to: (a) the fatigue susceptibility parameter $\bar{\alpha}_0$, and (b) the endurance parameter $\alpha_e$. Calculations obtained using \texttt{AT1}, $n=1$, $\kappa=0.5$ and the \texttt{No-tension} split.}
    \label{fig:SN_alphTE}
\end{figure}

Subsequently, a parametric study is conducted to investigate the influence of the fatigue model/material parameters. The calculations evaluating the sensitivity to $\bar{\alpha}_0$ and $\alpha_e$ are respectively shown in Fig. \ref{fig:SN_alphTE}a and Fig. \ref{fig:SN_alphTE}b, in terms of the remote stress amplitude versus the number of cycles to failure (S–N curves). The \texttt{AT1} model is used, the stress amplitude is normalised by the material strength, and the arrows correspond to the so-called fatigue runout phenomenon - samples that do not fail in the duration of the test. First, as can be seen in Fig. \ref{fig:SN_alphTE}a, the results reveal a longer fatigue life for higher values of $\bar{\alpha}_0$, in agreement with expectations. Second, Fig. \ref{fig:SN_alphTE}b showcases how decreasing the threshold parameter $\alpha_e$ leads to a decrease in the stress amplitude at which the fatigue life is practically infinite (the endurance limit). For both $\bar{\alpha}_0$ and $\alpha_e$, changes in their values do not lead to noticeable variations in the slope of the S-N curves.

\begin{figure}[hbt!]
\begin{subfigure}[t]{0.49\textwidth}
        \centering
    \includegraphics[width=1\linewidth,trim={0 0 0 0},clip]{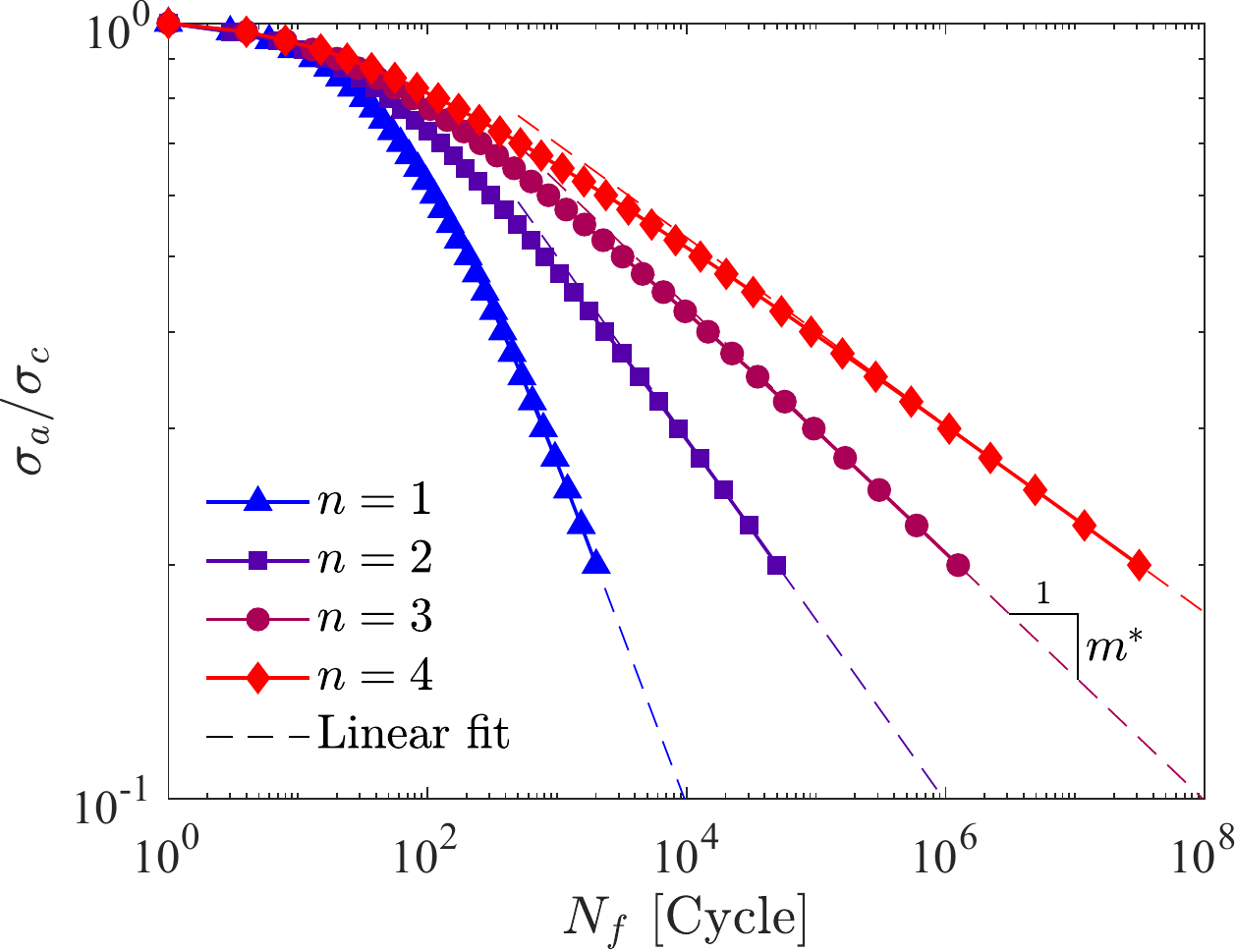}
    \caption{}
    \end{subfigure}
        \begin{subfigure}[t]{0.49\textwidth}
        \centering
    \includegraphics[width=0.96\linewidth,trim={0 0 0 0},clip]{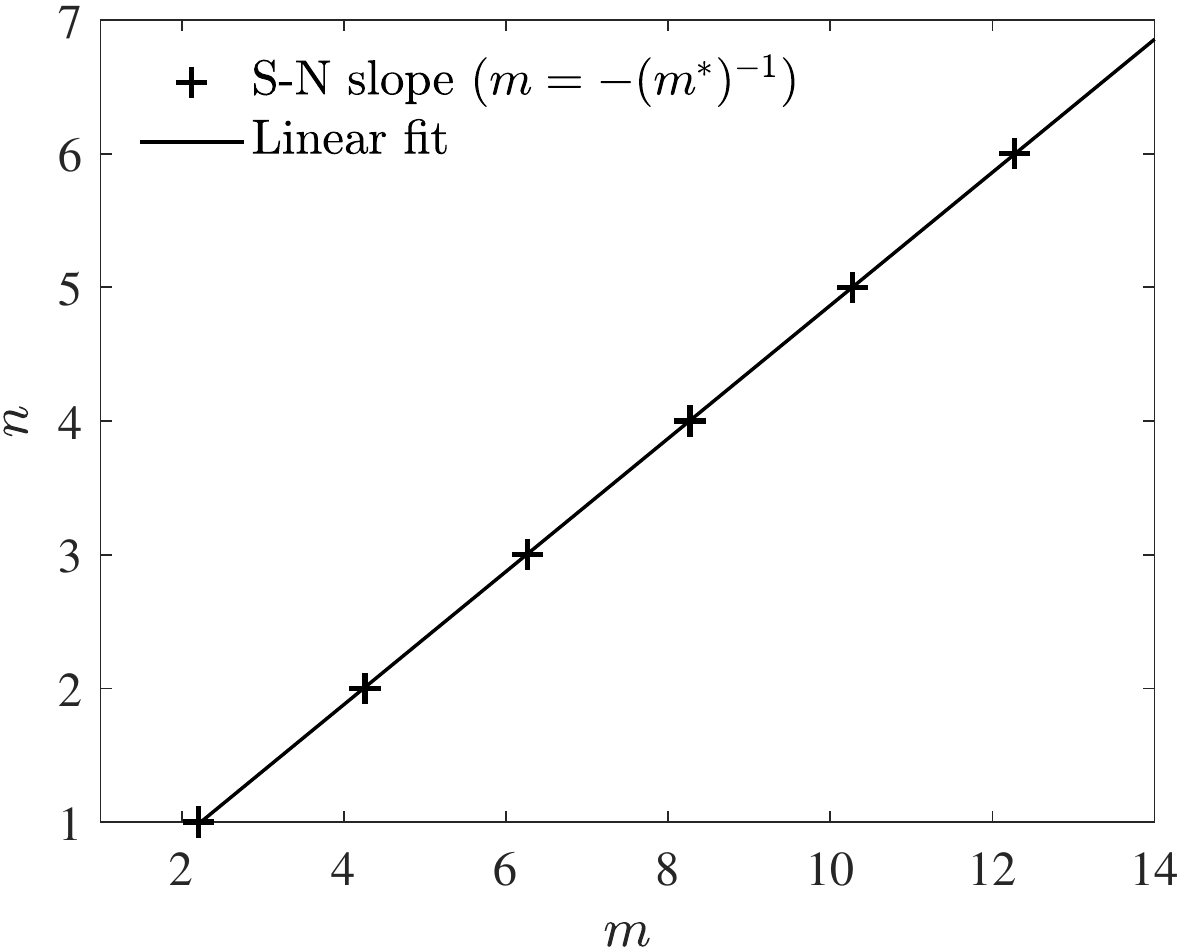}
    \caption{}
    \end{subfigure}
    \caption{Parametric study. Sensitivity to the power exponent $n$ in terms of the (a) S-N curve behaviour and (b) its linear behaviour with the S-N slope. Calculations obtained using \texttt{AT1}, $\kappa=0.5$ and the \texttt{No-tension} split.}
    \label{fig:SN_nm}
\end{figure}

Finally, the parametric study concludes with the investigation of the role of the power exponent $n$. The results are shown in Fig. \ref{fig:SN_nm}. The S-N curves show a clear dependence on the magnitude of $n$ (see Fig. \ref{fig:SN_nm}a), with larger $n$ values delivering fatigue responses that are more susceptible to changes in the stress amplitude. In other words, this parameter $n$ provides additional modelling flexibility and enables capturing the S-N curve slope $m^*$ of any material. As shown in Fig. \ref{fig:SN_nm}b, there exists a linear relationship between $n$ and $m$. Based on this finding we list in Table \ref{tab:nm}, for different phase field models and fatigue degradation functions, the coefficients of this linear relationship,
\begin{equation}\label{Eq:MvsNrelation}
    n= C_1 m + C_2
\end{equation}
where $m=-\left( m^*\right)^{-1}$. It is also worth noticing that, for higher stress amplitudes, the S-N curve deviates from such linear behaviour, demonstrating a damage-driven failure, as also reported by Carrara et al. \cite{Carrara2020}.\\

\begin{table}[hbt!]
\centering
\begin{tabular}{l|cc|cc|cc|}
\cline{2-7}
\multicolumn{1}{c|}{}     & \multicolumn{2}{c|}{$f_0$}         & \multicolumn{2}{c|}{$f_1$}         & \multicolumn{2}{c|}{$f_2$}         \\ \cline{2-7} 
\multicolumn{1}{c|}{}     & \multicolumn{1}{c}{$C_1$} & $C_2$ & \multicolumn{1}{c}{$C_1$} & $C_2$ & \multicolumn{1}{c}{$C_1$} & $C_2$ \\ \hline
\multicolumn{1}{|l|}{AT1} & \multicolumn{1}{c}{$0.50$}     & $-0.56$     & \multicolumn{1}{c}{$0.50$}     & $-0.63$     & \multicolumn{1}{c}{$0.50$}  & $-0.13$  \\ \hline
\multicolumn{1}{|l|}{AT2} & \multicolumn{1}{c}{$0.50$}     & $-0.55$     & \multicolumn{1}{c}{$0.49$}     & $-0.61$     & \multicolumn{1}{c}{$0.49$}     & $-0.12$     \\ \hline
\end{tabular}
\caption{Coefficients for the linear relationship between the power exponent $n$ and the S-N slope, see Eq. (\ref{Eq:MvsNrelation}).}
\label{tab:nm}
\end{table}
\subsubsection{Load ratio effect}

\begin{figure}[hbt!]
\begin{subfigure}[t]{0.49\textwidth}
        \centering
    \includegraphics[width=1\linewidth,trim={0 0 0 0},clip]{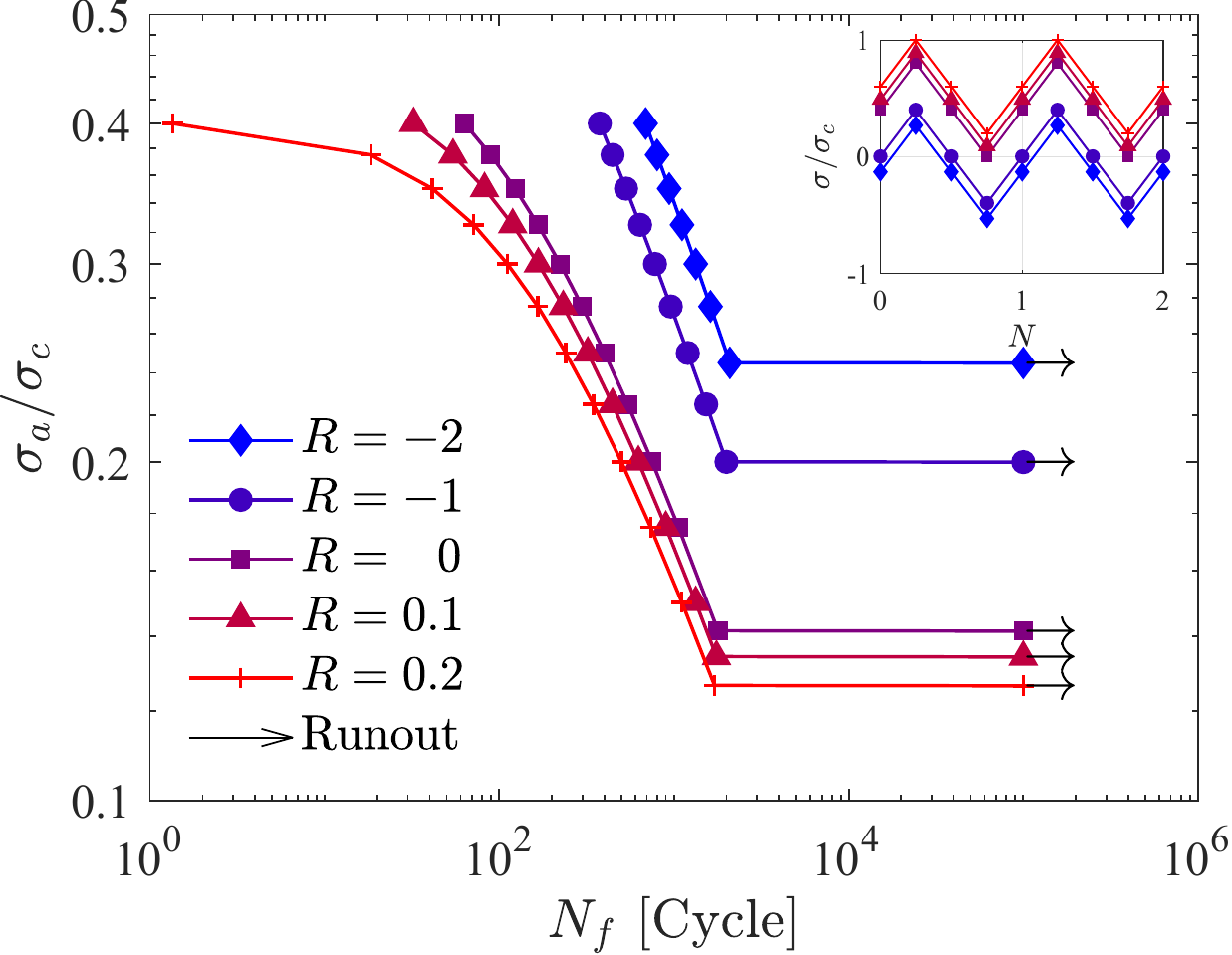}
    \caption{}
    \end{subfigure}
    \begin{subfigure}[t]{0.49\textwidth}
        \centering
    \includegraphics[width=1\linewidth,trim={0 0 0 0},clip]{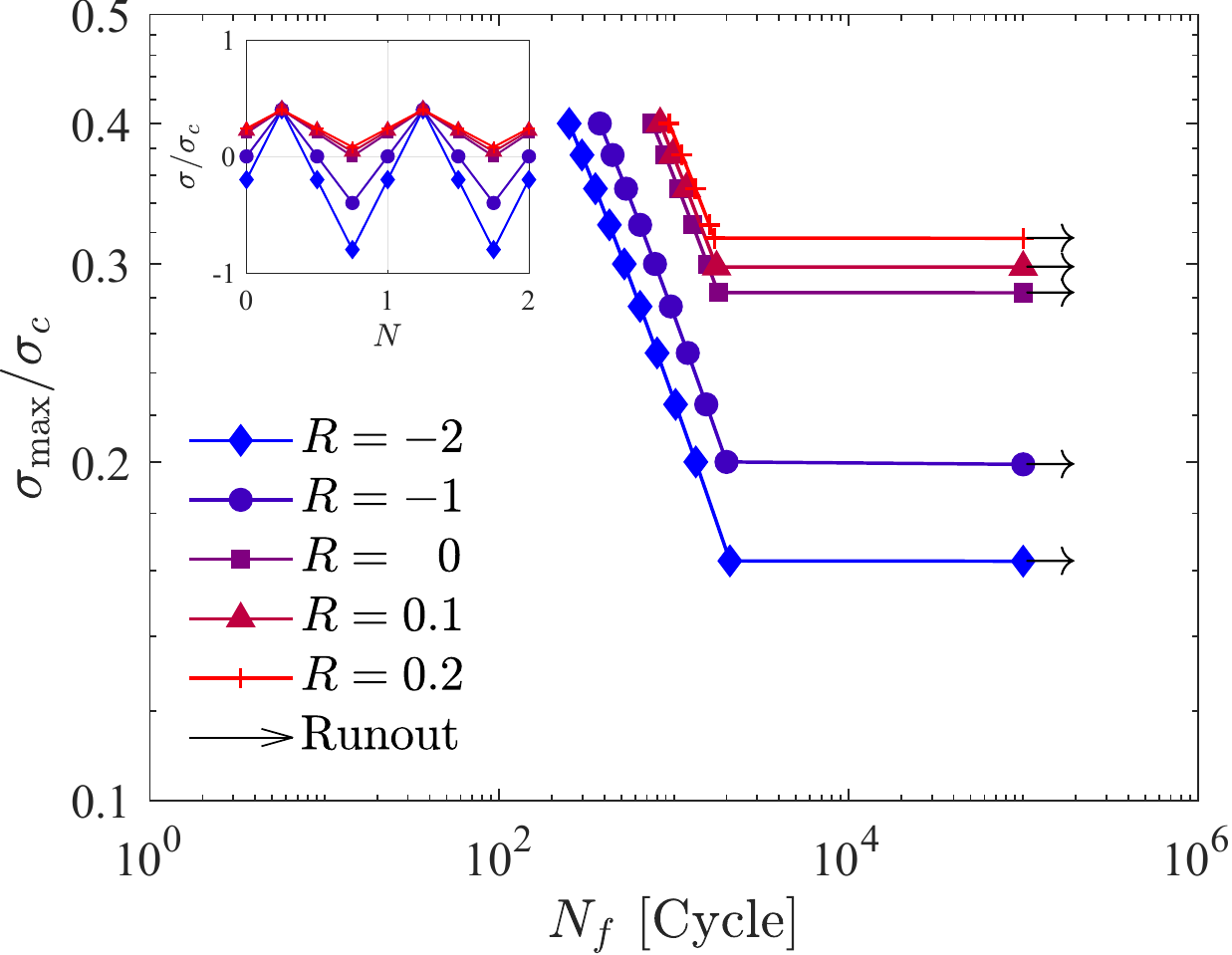}
    \caption{}
    \end{subfigure}
    \caption{Load ratio ($R$) effect, predictions obtained with (a) a fixed stress amplitude $\sigma_a$, and (b) a fixed maximum stress $\sigma_{\mathrm{max}}$. The subplots illustrate the loading conditions, for the specific cases of $\sigma_a/\sigma_c=0.4$ and $\sigma_{\mathrm{max}}/\sigma_c=0.4$. Calculations obtained using \texttt{AT1}, $n=1$, $\kappa=0.5$ and the \texttt{No-tension} split.}
    \label{fig:SN_theory_R}
\end{figure}

We shall now investigate the ability of the proposed model to capture the mean stress effect on S-N curve behaviour. To this end, two load-controlled scenarios are considered: (i) a varying $R$ for a fixed stress amplitude $\sigma_a$, and (ii) a varying $R$ for a fixed maximum stress $\sigma_{\mathrm{max}}$. These loading scenarios are of particular interest because experimental observations report opposite trends in terms of $R$ vs number of cycles behaviour, with fixed $\sigma_a$ experiments showing a longer fatigue life for decreasing $R$ while the opposite is observed for fixed $\sigma_{\mathrm{max}}$ tests \cite{MIL1998,Dowling2009}. The results obtained are given in Fig. \ref{fig:SN_theory_R}, together with a subplot depicting the loading conditions for the cases of $\sigma_a/\sigma_c=0.4$ and $\sigma_{\mathrm{max}}/\sigma_c=0.4$. A significant influence of the load ratio $R$ on the fatigue life and the endurance limit is observed, for both loading scenarios. Consider first the fixed stress amplitude case, Fig. \ref{fig:SN_theory_R}a. For a given $\sigma_a$, the fatigue life decreases significantly with increasing the load ratio $R$, in agreement with experimental observations \cite{Dowling2009}. It can also be observed that, for higher load ratios, the S-N curve exhibits non-linear behaviour with a notable drop in the fatigue life. This can be explained by the fact that, for higher load ratios, the maximum value of the cyclic stress observed in the subplot reaches the material critical strength $\sigma_c$, suggesting that the failure is governed by static damage rather than fatigue (see also Fig. \ref{fig:SN_nm}a). Next, consider the constant $\sigma_{\mathrm{max}}$ results in Fig. \ref{fig:SN_theory_R}b. Contrarily to what is observed in the constant $\sigma_a$ case, and in agreement with experiments (see Ref. \cite{MIL1998} and the experimental comparison below), fatigue lives increase with increasing $R$. Thus, the generalised model presented is able to adequately capture the sensitivity to the load ratio $R$ under both constant stress amplitude and constant maximum stress. 

\subsection{Comparison with experimental S-N curves}

\begin{figure}[hbt!]
\centering
\includegraphics[width=0.9\linewidth]{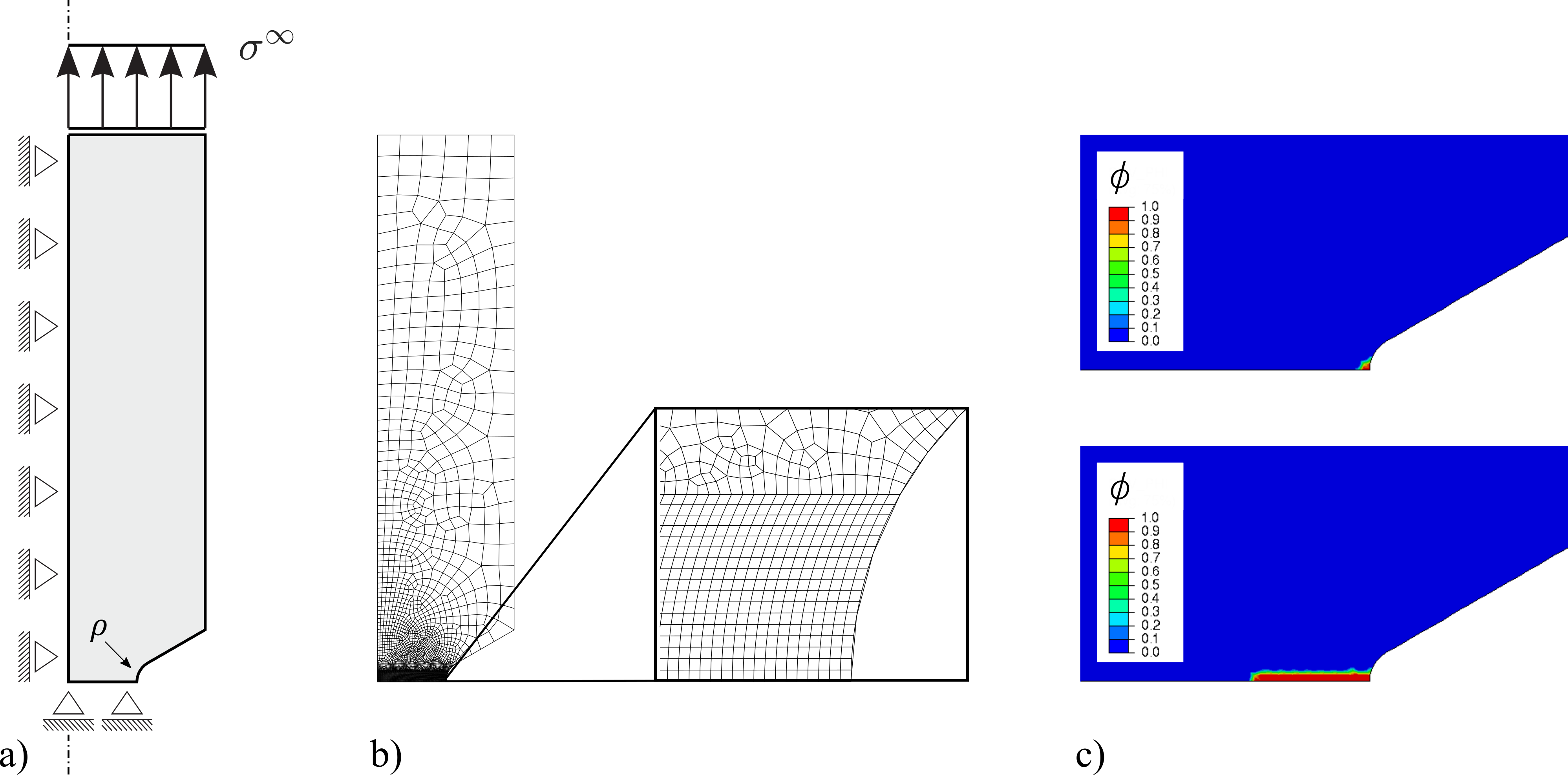}
\caption{Notched cylindrical bar ($60^{\circ}$ V-Groove): (a) geometry and boundary conditions, (b) finite element mesh, including a detailed view of the mesh ahead of the notch tip, and (c) representative phase field contours showing crack initiation and growth (up to the unstable failure event) for 300M steel with $K_t=5$ and $\sigma_{\text{max}}^{\text{nom}}=300$ MPa.}
\label{fig:Bar_Geometry}
\end{figure}


We proceed now to compare model predictions with S–N curves obtained from uniaxial tension-compression fatigue experiments on cylindrical bars, considering both smooth and notched samples. The experimental data are taken from Ref. \cite{MIL1998} and correspond to two types of low-alloy steels, an AISI 4340 steel with tensile strength of 1,793 MPa, and a 300M steel with tensile strength of 2,000 MPa. The experiments were carried out in laboratory air under constant maximum stress amplitudes at various stress ratios $R$. As is common among steels, both materials are assumed to have a Young's modulus of $E=210$ GPa and a Poisson's ratio of $\nu=0.3$. The toughnesses values are taken to be equal to $G_c=20$ kJ/m$^2$ and $G_c=13$ kJ/m$^2$ for AISI 4340 and 300M, respectively, based on plane strain fracture toughness measurements reported in Ref. \cite{stephens2000}. Results for the unnotched samples can be obtained semi-analytically, considering the homogeneous solution to (\ref{eq:StrongFormPhi}). For the notched samples, finite element calculations are conducted, where axial symmetry is exploited to consider only one planar section of the sample. In addition, only the upper half of the domain is modelled due to vertical symmetry (see Fig. \ref{fig:Bar_Geometry}). The finite element domain is discretised using 4-node bilinear axisymmetric quadrilateral elements with full integration, with the mesh being refined ahead of the notch tip, where the characteristic element size is 10 times smaller than the phase field length scale $\ell$ (see Fig. \ref{fig:Bar_Geometry}b). Under 1D conditions, the length scale and the strength are related via (\ref{eq:sigmac}), and this relation renders magnitudes of $\ell=0.318$ mm and $\ell=0.315$ mm for AISI 4340 and 300M, respectively. For the 300M notched samples, the notch radii magnitudes considered are $\rho=1.016$, $0.368$, and $0.107$ mm, with the bar gross diameter being $D=12.7$ mm and the net diameter $d=6.35$ mm. From these, the following stress concentration factors (SCF) are obtained: $K_t=2$, $3$, and $5$. For the case of AISI 4340, the notch radii magnitudes read $\rho=0.762$ and $0.254$ mm. The following diameters are considered: $D=7.62$ mm, $D=6.86$ mm, and $d=5.59$ mm, which correspond to SCF values of $K_t=2$ and $3$. The samples are subjected to a piece-wise cyclic linear force-controlled loading with a load ratio of $R=-1$. The endurance limit is estimated from the S-N curve itself at the stress level below which infinite life is expected; the magnitudes of $\sigma_e=530$ MPa and $\sigma_e=650$ MPa are assumed for AISI 4340 and 300M, respectively. The slope of the S-N curve and its intercept with the $\log N$ axis are, respectively, linked to the fatigue parameters $n$ and $\bar{\alpha}_0$ (see Fig. \ref{fig:SN_nm} and \ref{Sec:appendixB}). Thanks to this feature, the fatigue parameters $n$ and $\bar{\alpha}_0$ can now be estimated so as to provide the best fit to the experiments of unnotched (smooth) samples subjected to fully-reversed cyclic loading ($R=-1$); the magnitudes of $\bar{\alpha}_0=5.0\times10^{-4}$, $n=10$ and $\bar{\alpha}_0=1.7\times10^{1}$, $n=6$, respectively, provided a good agreement with the experiments on AISI 4340 and 300M. Accordingly, any other effects (e.g. the role of notch radius or the sensitivity to the loading ratio) are predicted as a natural outcome of the model, without any additional fitting.

\begin{figure}[hbt!]
\begin{subfigure}[t]{0.49\textwidth}
        \centering
    \includegraphics[width=1\linewidth,trim={0 0 0 0},clip]{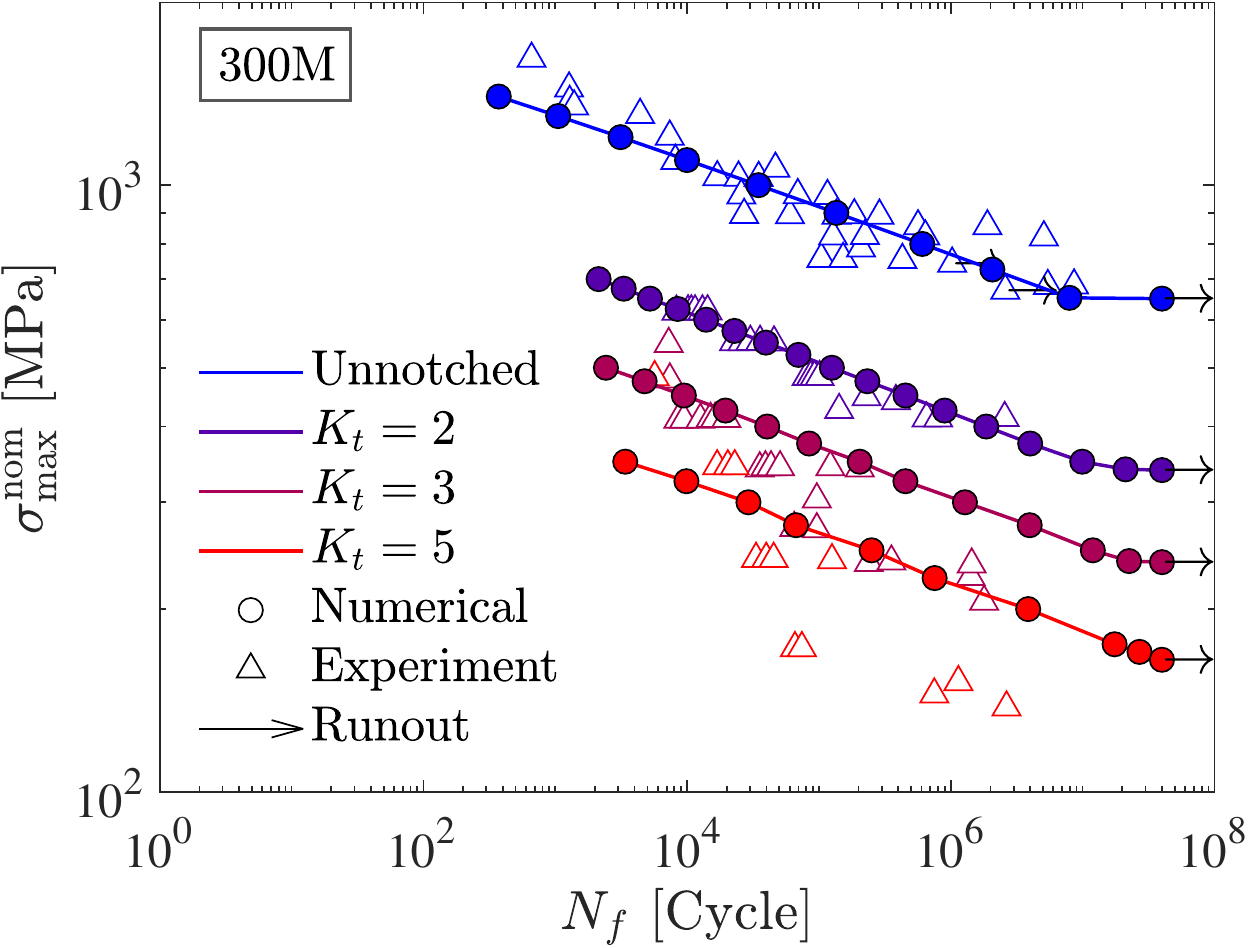}
    \caption{}
    \end{subfigure}
        \begin{subfigure}[t]{0.49\textwidth}
        \centering
    \includegraphics[width=1\linewidth,trim={0 0 0 0},clip]{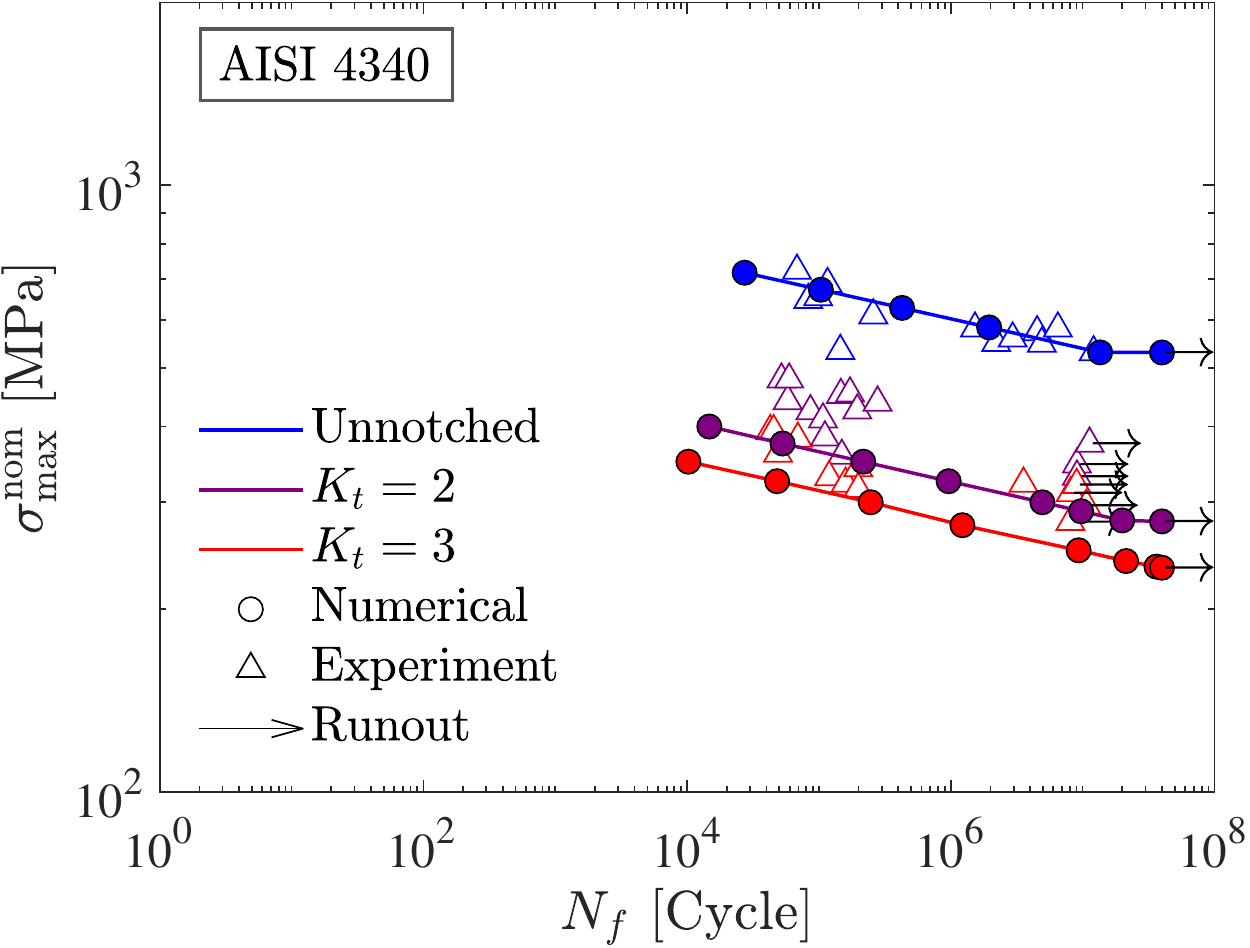}
    \caption{}
    \end{subfigure}
    \caption{Experimental validation. Numerical and experimental \cite{MIL1998} S–N curves obtained from smooth and notched cylindrical bars for two types of steel: (a) 300M, and (b) AISI 4340. The model is shown to be able to predict the role of stress raisers (as quantified by the stress concentration factor $K_t$) in reducing fatigue lives.}
    \label{fig:SN_exp_Kt}
\end{figure}

The experimental and numerical results obtained are shown in Fig. \ref{fig:SN_exp_Kt}. It can be seen that the \emph{Virtual} S-N curves predicted are in good agreement with the measured data. In both experiments and simulations, the results demonstrate a strong sensitivity to the notch radius, with the fatigue life decreasing by reducing the radius. Smaller radii result in higher stress concentrations at the notch tip, leading to an earlier initiation of the fatigue crack, as expected. It is also worth noting that the agreement with experiments of 300M steel becomes less satisfactory at smaller notch radii ($K_t=5$), as the slope of the experimental S-N curve exhibits a change. This change in slope for the case of $K_t=5$ could be related to plastic phenomena such as the reverse yielding effect \cite{McClung1991}.

Also, as shown for the AISI 4340 experiments, the model readily captures the influence of stress concentrations on the endurance limit. Overall, the model is shown to be able to reliably predict the fatigue lives and endurance limit of samples containing different notches (stress concentrators) without the need for fitting.

\begin{figure}[hbt!]
\begin{subfigure}[t]{0.49\textwidth}
        \centering
    \includegraphics[width=1\linewidth,trim={0 0 0 0},clip]{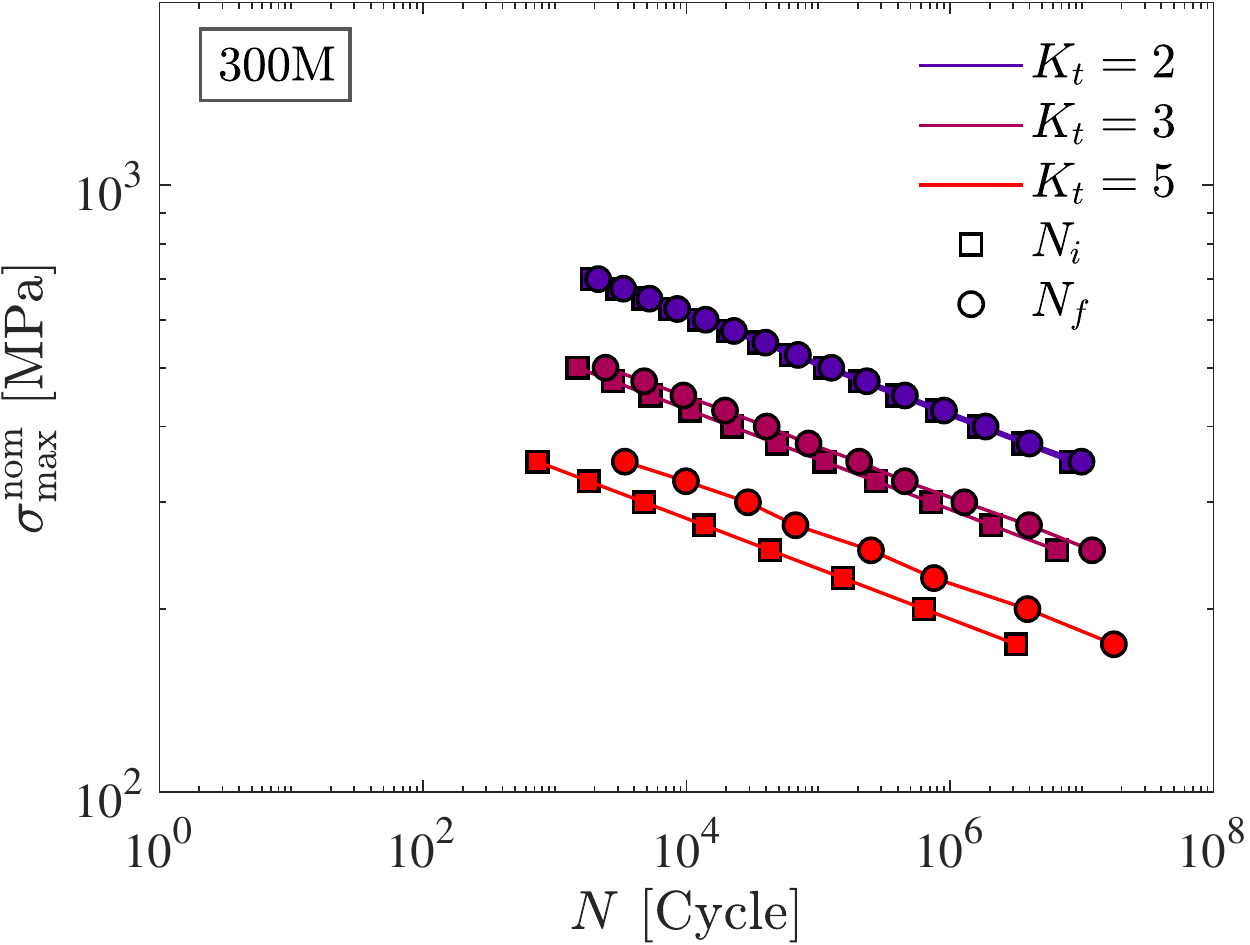}
    \caption{}
    \end{subfigure}
        \begin{subfigure}[t]{0.49\textwidth}
        \centering
    \includegraphics[width=1\linewidth,trim={0 0 0 0},clip]{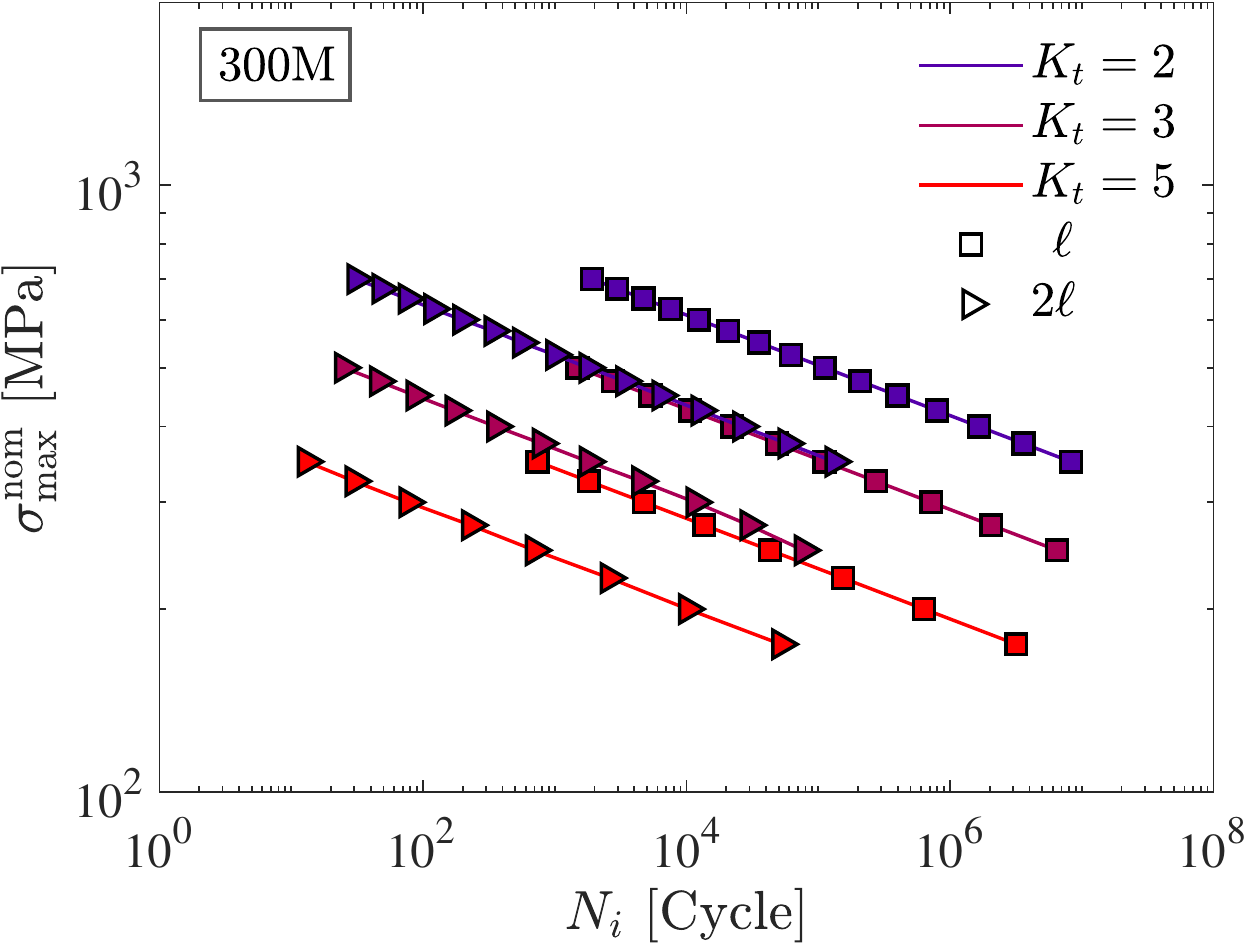}
    \caption{}
    \end{subfigure}
    \caption{S–N curves behaviour predicted for notched cylindrical bars: (a) comparison between the number of cycles for crack initiation ($N_i$) and the number of cycles to failure ($N_f$), and (b) interplay between the phase field length scale $\ell$ and the notch radius $\rho$. Results are obtained for the parameters relevant to 300M steel.}
    \label{fig:SN_exp4_Kt_extra}
\end{figure}

Building upon the 300M results, we use the model to gain further insight into the material fatigue behaviour. First, as shown in Fig. \ref{fig:SN_exp4_Kt_extra}a, the the number of cycles to initiation and failure is plotted as a function of maximum nominal stress $\sigma_{\text{max}}^{\text{nom}}$ and the stress concentration factor $K_t$. The results reveal that the differences between crack nucleation and final failure increase as the notch becomes sharper. This is the result of the stronger localisation of stress, strain and damage in sharper defects. Then, we investigate the interplay between length scales by varying the phase field length scale parameter $\ell$, for a fixed notch radius $\rho$ - see Fig. \ref{fig:SN_exp4_Kt_extra}b. Specifically, we choose to consider a value of $\ell$ twice as high (i.e., $2\ell=0.63$ mm). The results show that the fatigue resistance decreases with increasing $\ell$. This is in agreement with expectations as, according to Eq. (\ref{eq:sigmac}), a higher value of $\ell$ will lead to a decrease in material strength and thus a shorter time to crack nucleation. It is worth noting that the values of $\ell$ considered are on the order of the notch radius. However, the results do not scale with $\ell/\rho$, suggesting the influence of other length scales in the problem. This can be seen by considering the results for $K_t=2$ and $2\ell$ and the ones for $K_t=3$ and $\ell$, which respectively give $\rho/\ell=1.168$ and $\rho/\ell=1.613$, yet appear to fall on top of each other. A dimensional analysis could be carried out to establish the calculations needed to understand the interplay between the various length scales of the problem.

\begin{figure}[hbt!]
\begin{subfigure}[t]{0.49\textwidth}
        \centering
    \includegraphics[width=1\linewidth,trim={0 0 0 0},clip]{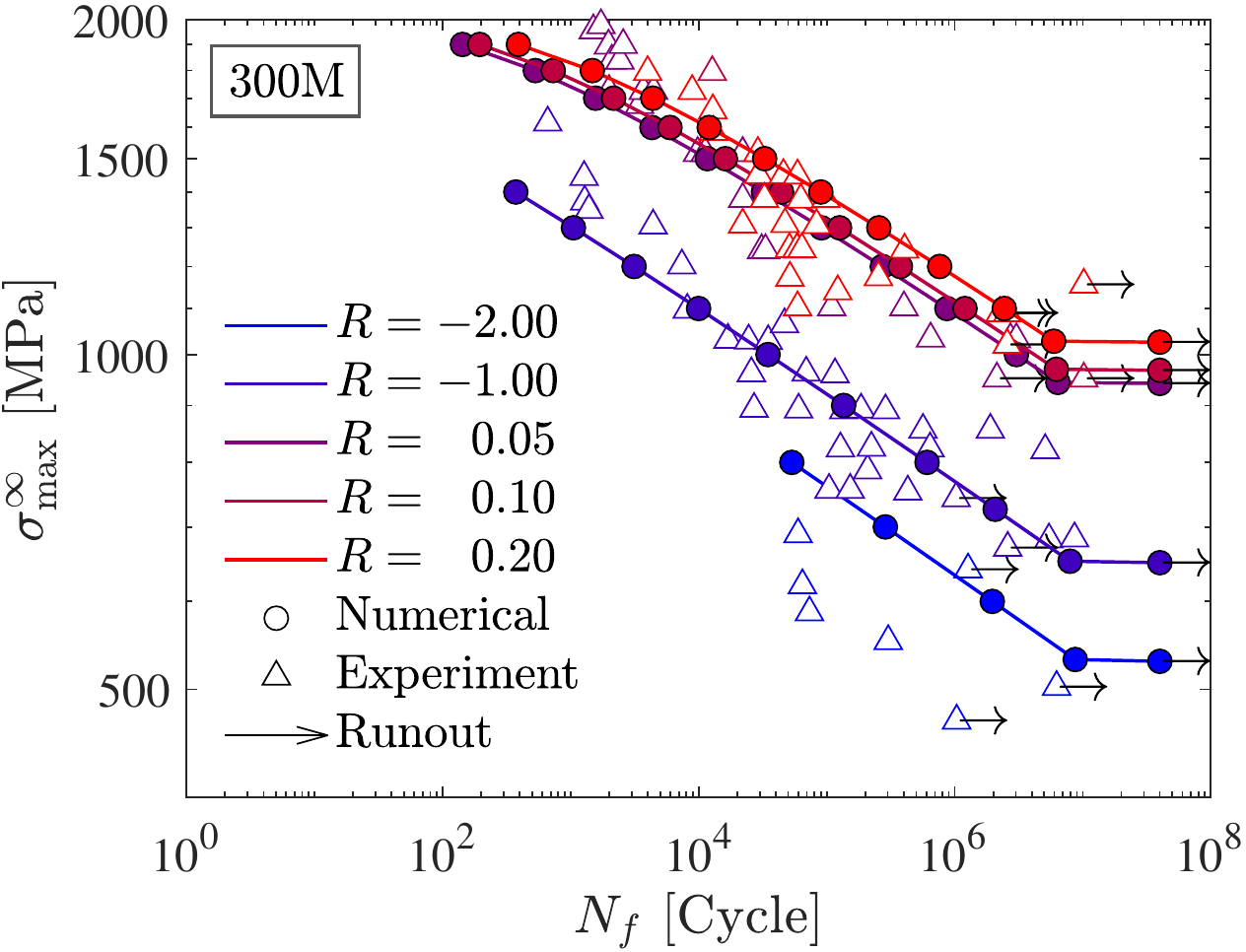}
    \caption{}
    \end{subfigure}
        \begin{subfigure}[t]{0.49\textwidth}
        \centering
    \includegraphics[width=1\linewidth,trim={0 0 0 0},clip]{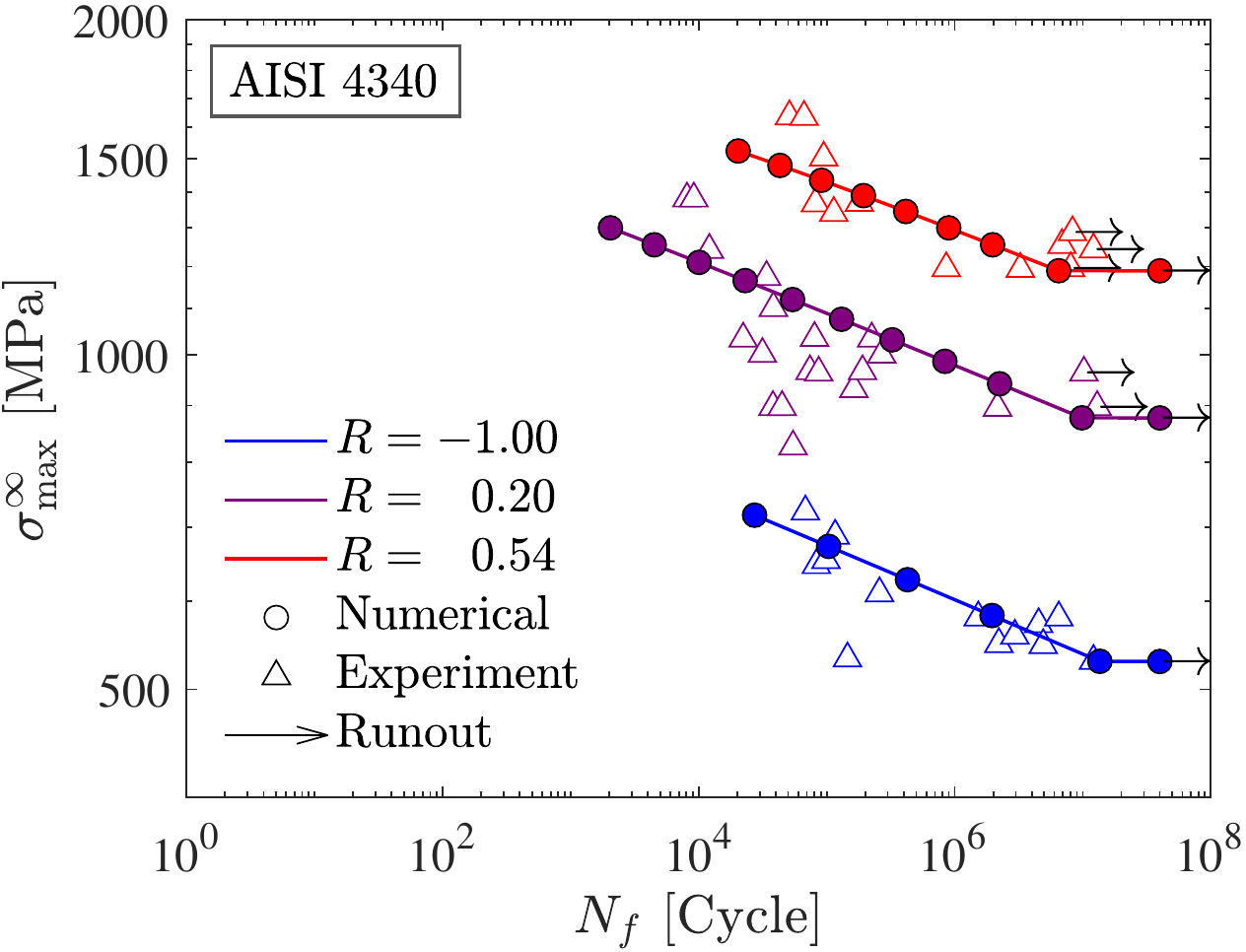}
    \caption{}
    \end{subfigure}
    \caption{Experimental validation. Numerical and experimental \cite{MIL1998} S–N curves obtained from smooth cylindrical bars at various load ratios $R$ for two types of steel: (a) 300M, and (b) AISI 4340. The model is shown to be able to predict the role of the load ratio in varying the fatigue resistance of the material.}
    \label{fig:SN_exp_R}
\end{figure}

Finally, we validate model predictions of the load ratio effect against experiments on the two steels considered above (300M and AISI 4340). The model parameters are those considered before. In this regard, it should be noted that $\kappa$ is taken to be equal to 0.55 and 0.5 for AISI 4340 and 300M, respectively, based on estimations reported in \cite{Dowling2009}. However, similar results would be obtained considering the simpler Smith-Watson-Topper (SWT) relationship, thus eliminating the need for this parameter altogether. The numerical predictions are shown together with experimental data in Fig. \ref{fig:SN_exp_R}. Both numerical and experimental data reveal the same qualitative trend: for a fixed $\sigma_{\text{max}}^\infty$, the number of cycles to failure $N_f$ increases with increasing load ratio $R$. Moreover, for both 300M and AISI 4340 materials, the model delivers a good quantitative agreement with experiments, demonstrating the ability of the model to successfully predict the mean stress effect. Some differences are observed for the specific case of $R=-2$ and 300M, where the samples are under compression for the majority of their fatigue lives and the experimental scatter is notable. 

\section{Conclusions}
\label{Sec:Conclusions}
We have formulated a generalised phase field formulation for modelling high-cycle fatigue behavior in metallic materials. The modelling framework presented encompasses the two main phase field damage models (\texttt{AT1} and \texttt{AT2}), different fatigue degradation functions, and a new accumulation approach that significantly accelerates calculations and allows modelling: (i) different S-N curve slopes, (ii) the fatigue endurance limit, and (iii) the mean stress effect (load/stress ratio). The theoretical framework presented is numerically implemented using the finite element method and the resulting system of equations is solved in a monolithic manner, by using a robust and efficient quasi-Newton (BFGS) algorithm. Total-life analyses are conducted to investigate the performance of the modelling abilities of the proposed framework. The influence on fatigue damage accumulation of various strain energy decomposition approaches (volumetric/deviatoric, spectral, no-tension) is investigated. Also, \emph{Virtual} S–N curves are obtained for various stress/load ratios and for both notched and smooth samples. Key findings include:
\begin{itemize}    
    \item The model adequately captures the sensitivity of fatigue life to the presence of stress raisers (such as notches), with both fatigue life and endurance limit decreasing with increasing stress concentration.
    \item The mean stress effect (load ratio, $R$) on the fatigue response is adequately captured. In agreement with experimental observations, the model predicts an increase in fatigue life and endurance limit with decreasing $R$ for a fixed stress amplitude $\sigma_a$, while the opposite is true for a fixed maximum stress $\sigma_{\text{max}}$.
    \item The agreement with experiments is both qualitative and quantitative, with the model providing a good agreement with fatigue lives and endurance limit data for 300M and AISI 4340 steels. Moreover, the role of stress raisers and load ratio on the fatigue response of these two materials is naturally captured.
\end{itemize}

The modelling framework presented provides a platform to efficiently predict the service lives of components undergoing high-cycle fatigue. Potential avenues for future work could be directed towards the development of a generalised model that could also consider low- and mid-cycle fatigue, plasticity effects and Paris law behaviour.

\section{Acknowledgements}
\label{Acknowledge of funding}
A. Golahmar acknowledges financial support from Vattenfall Vindkraft A/S and Innovation Fund Denmark (grant 0153-00018B). E. Mart\'{\i}nez-Pa\~neda acknowledges financial support from UKRI's Future Leaders Fellowship programme [grant MR/V024124/1].
\begin{appendix}
\section{Comparison with existing phase field fatigue models}
\label{Sec:appendixA}
\begin{figure}[H]
    \centering
    \includegraphics[width=0.49\linewidth]{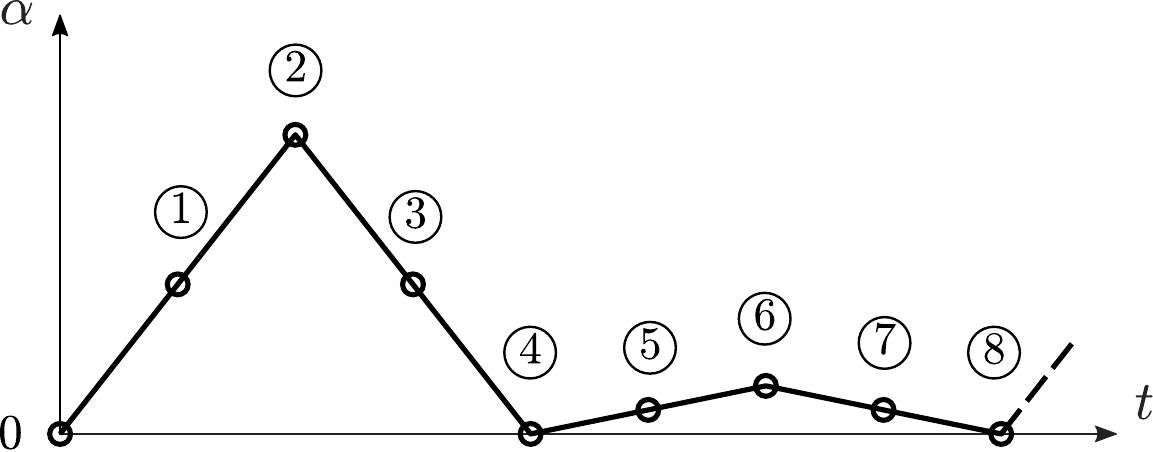}
    \caption{Schematic variation of $\alpha$ for a fully-reversed cyclic loaded ($R=-1$) bar using the \texttt{Spectral} split.}
    \label{fig:cyclic_alpha}
\end{figure}

Considering the following approximation for $\Delta \bar{\alpha}$ as suggested in Ref. \cite[Eq. (45)]{Carrara2020}
\begin{equation}
    \Delta \bar{\alpha}= \left|\alpha_{n+1}-\alpha_n\right| H\left(\frac{\alpha_{n+1}-\alpha_n}{t_{n+1}-t_n}\right)
    \label{eq:alphB_Carrara}
\end{equation}
where the subscripts ${ }_n$ and ${ }_{n+1}$ refer to the time increments $t=t_n$ and $t=t_{n+1}$, respectively. The Heaviside function $H(\square)=0$ when $\Delta \alpha/\Delta t<0$ (unloading). We now proceed to calculate the total increase of the fatigue history variable $\Bar{\alpha}$ considering a fully-reversed cyclic loading ($R=-1$) of a bar, using the \texttt{Spectral} decomposition split (\ref{eq:decMiehe}) and 8 load steps per cycle (see Fig. \ref{fig:cyclic_alpha})
\begin{equation}
\begin{split}
     \text{Inc. 0 - 4:}\quad& \Bar{\alpha}_4=\Bar{\alpha}_3=\alpha_2=\Bar{\alpha}_1+|\alpha_2-\alpha_1|=\alpha_2\\[0.3mm]
     \text{Inc. 4 - 8:}\quad&\Bar{\alpha}_8=\Bar{\alpha}_7=\Bar{\alpha}_6=\Bar{\alpha}_5+|\alpha_6-\alpha_5|=\underbrace{\alpha_2+\alpha_6}_{\alpha_\mathrm{max}+\alpha_\mathrm{min}}\\[0.3mm]
\end{split}    
\end{equation}
which demonstrates that the accumulation of fatigue damage at the end of each cycle can be described by the values of $\alpha$ obtained at the peak $\alpha_2$ and valley $\alpha_6$ during one reversal (see also Fig. \ref{fig:cyclic_stress}). Thus, Eq. (\ref{eq:alphB_Carrara}), could be reformulated as
\begin{equation}
    \Delta \bar{\alpha}=\frac{\alpha_{\text{max}}^n - \mathrm{sgn}(R)\, \alpha_{\text{min}}^n}{\alpha_n^n}
    \label{eq:dalphB_1}
\end{equation}
where the stress ratio $R$ and its sign $\mathrm{sgn}(R)$ can be computed for each material point, on the fly, within each cycle. For a specific choice of $n=1$ and $\alpha_n=1$, Eq. (\ref{eq:dalphB_1}) recovers Eq. (\ref{eq:alphB_Carrara}), at the end of each cycle, for any arbitrary stress ratio $R$ when using the \texttt{No-tension} split (and for $R \geq 0$ and $R=-1$ when using the other splits). In addition, for constant amplitude cases, one could accelerate the calculation of $\Delta \Bar{\alpha}$ by using only one increment per cycle and applying a constant (representative) load with the maximum value of the amplitude as its magnitude. Thus, Eq. (\ref{eq:dalphB_1}) can be altered as
\begin{equation}
    \Delta \bar{\alpha}=\left(\frac{\alpha_{\text{max}}}{\alpha_n}\right)^n\left(1-\mathrm{sgn}(R){\left|R\right|}^{2 n}\right)
    \label{eq:dalphB_2}
\end{equation}
which yields identical analytical results to Eqs. (\ref{eq:dalphB_1}) and (\ref{eq:alphB_Carrara}) for a fixed stress ratio $R \geq 0 $ when using the \texttt{No-tension} and \texttt{Volumetric-deviatoric} splits. Finally, for a specific choice of $n=1$, $\alpha_n=1$, $R=-1$ and $\alpha_e=0$, our new accumulation approach (\ref{eq:dalphB_3}) recovers analytically Eq. (\ref{eq:alphB_Carrara}) when using the \texttt{No-tension} split.

\section{Estimation of the fatigue material parameter $\bar{\alpha}_0$}
\label{Sec:appendixB}
Considering a typical S-N curve obtained from a fatigue experiment and described mathematically by the Basquin relationship $\sigma_{\square}=C^*\left(N_{\square}\right)^{m^*}$ where $(N_\square,\sigma_{\square})$ corresponds to the data set $\square$ of the fitted curve. As illustrated in Fig. \ref{fig:SN_nm}, the slope of the S-N curve $m^*$ is linked to the power exponent $n$, with the fitting parameters presented in Table \ref{tab:nm} for different choices of phase field damage model and fatigue degradation function. We now proceed to estimate the fatigue material parameter $\bar{\alpha}_0$, by considering the homogeneous solution to (\ref{eq:StrongFormPhi}) and assuming an \emph{undamaged} strain energy density for $\alpha=\psi_0^+(\varepsilon)$. Then, considering the \texttt{AT1} damage model, the $f_2$ fatigue degradation function, and the fact that $f(\bar{\alpha})=1$ for $\sigma=\sigma_c$ (static loading), then
\begin{equation}
    \left(\dfrac{\sigma_\square}{\sigma_c}\right)^2=f(\bar{\alpha})
    =\left(1-\dfrac{N_\square}{\bar{\alpha}_0}\left(\dfrac{\alpha_\mathrm{max}}{\alpha_c}\right)^n\right)^2
    =\left(1-\dfrac{N_\square}{\bar{\alpha}_0}\left(\dfrac{\sigma_\square}{\sigma_c}\right)^{2 n}\right)^2 
\end{equation}
\noindent which results in
\begin{equation}
        \bar{\alpha}_0=\dfrac{N_\square \left(\dfrac{\sigma_\square}{\sigma_c}\right)^{2 n}}{1-\left(\dfrac{\sigma_\square}{\sigma_c}\right)}
\end{equation}
for which a good estimation can be obtained by using low stress magnitudes for $\sigma_\square$ (and consequently higher fatigue lives for $N_\square$), where the S-N curve is not deviating from linearity. 

\end{appendix}


\begin{thebibliography}{10}
\expandafter\ifx\csname url\endcsname\relax
  \def\url#1{\texttt{#1}}\fi
\expandafter\ifx\csname urlprefix\endcsname\relax\def\urlprefix{URL }\fi
\expandafter\ifx\csname href\endcsname\relax
  \def\href#1#2{#2} \def\path#1{#1}\fi

\bibitem{stephens2000}
R.~I. Stephens, A.~Fatemi, R.~R. Stephens, H.~O. Fuchs, Metal Fatigue in
  Engineering, 2nd Edition, John Wiley \& Sons, 2000.

\bibitem{Suresh1998}
S.~Suresh, Fatigue of Materials, 2nd Edition, Cambridge University Press, 1998.

\bibitem{Wohler1870}
A.~Wöhler, Über die festigkeitsversuche mit eisen und stahl, Ernst \& Korn,
  1870.

\bibitem{Griffith1920}
A.~A. Griffith, The phenomena of rupture and flow in solids, Philosophical
  Transactions A, 221 (1920) 163--198.

\bibitem{Francfort1998}
G.~A. Francfort, J.-J. Marigo, Revisiting brittle fracture as an energy
  minimization problem, Journal of the Mechanics and Physics of Solids 46
  (1998) 1319--1342.

\bibitem{Bourdin2008}
B.~Bourdin, G.~A. Francfort, J.~J. Marigo, The variational approach to
  fracture, Springer Netherlands, 2008.

\bibitem{Ambati2015b}
M.~Ambati, T.~Gerasimov, L.~D. Lorenzis, Phase-field modeling of ductile
  fracture, Computational Mechanics 55 (2015) 1017--1040.

\bibitem{Borden2016}
M.~J. Borden, T.~J.~R. Hughes, C.~M. Landis, A.~Anvari, I.~J. Lee, A
  phase-field formulation for fracture in ductile materials: Finite deformation
  balance law derivation, plastic degradation, and stress triaxiality effects,
  Computer Methods in Applied Mechanics and Engineering 312 (2016) 130--166.

\bibitem{Isfandbod2021}
M.~Isfandbod, E.~Martínez-Pañeda, A mechanism-based multi-trap phase field
  model for hydrogen assisted fracture, International Journal of Plasticity 144
  (2021) 103044.

\bibitem{Borden2012}
M.~J. Borden, C.~V. Verhoosel, M.~A. Scott, T.~J.~R. Hughes, C.~M. Landis, A
  phase-field description of dynamic brittle fracture, Computer Methods in
  Applied Mechanics and Engineering 217-220 (2012) 77--95.

\bibitem{Geelen2019}
R.~J.~M. Geelen, Y.~Liu, T.~Hu, M.~R. Tupek, J.~E. Dolbow, A phase-field
  formulation for dynamic cohesive fracture, Computer Methods in Applied
  Mechanics and Engineering 348 (2019) 680--711.

\bibitem{Molnar2020}
G.~Molnár, A.~Gravouil, R.~Seghir, J.~Réthoré, An open-source abaqus
  implementation of the phase-field method to study the effect of plasticity on
  the instantaneous fracture toughness in dynamic crack propagation, Computer
  Methods in Applied Mechanics and Engineering 365 (2020) 113004.

\bibitem{Alessi2019}
R.~Alessi, F.~Freddi, Failure and complex crack patterns in hybrid laminates: A
  phase-field approach, Composites Part B: Engineering 179 (2019) 107256.

\bibitem{Mandal2020}
T.~K. Mandal, V.~P. Nguyen, J.-Y. Wu, A length scale insensitive anisotropic
  phase field fracture model for hyperelastic composites, International Journal
  of Mechanical Sciences 188 (2020) 105941.

\bibitem{Quintanas-Corominas2020}
A.~Quintanas-Corominas, A.~Turon, J.~Reinoso, E.~Casoni, M.~Paggi, J.~A.
  Mayugo, A phase field approach enhanced with a cohesive zone model for
  modeling delamination induced by matrix cracking, Computer Methods in Applied
  Mechanics and Engineering 358 (2020) 112618.

\bibitem{Hirshikesh2019}
Hirshikesh, S.~Natarajan, R.~K. Annabattula, E.~Martínez-Pañeda, Phase field
  modelling of crack propagation in functionally graded materials, Composites
  Part B: Engineering 169 (2019) 239--248.

\bibitem{Kumar2021}
P.~K. A.~V. Kumar, A.~Dean, J.~Reinoso, P.~Lenarda, M.~Paggi, Phase field
  modeling of fracture in functionally graded materials: G -convergence and
  mechanical insight on the effect of grading, Thin-Walled Structures 159
  (2021) 107234.

\bibitem{Martinez-Paneda2018}
E.~Martínez-Pañeda, A.~Golahmar, C.~F. Niordson, A phase field formulation
  for hydrogen assisted cracking, Computer Methods in Applied Mechanics and
  Engineering 342 (2018) 742--761.

\bibitem{Duda2018}
F.~P. Duda, A.~Ciarbonetti, S.~Toro, A.~E. Huespe, A phase-field model for
  solute-assisted brittle fracture in elastic-plastic solids, International
  Journal of Plasticity 102 (2018) 16--40.

\bibitem{Wu2020a}
J.-Y. Wu, T.~K. Mandal, V.~P. Nguyen, A phase-field regularized cohesive zone
  model for hydrogen assisted cracking, Computer Methods in Applied Mechanics
  and Engineering 358 (2020) 112614.

\bibitem{Wu2020b}
J.-Y. Wu, V.~P. Nguyen, C.~T. Nguyen, D.~Sutula, S.~Sinaie, S.~Bordas,
  Phase-field modelling of fracture, Advances in Applied Mechanics 53 (2020)
  1--183.

\bibitem{Kristensen2021}
P.~K. Kristensen, C.~F. Niordson, E.~Martínez-Pañeda, An assessment of phase
  field fracture: crack initiation and growth, Philosophical Transactions of
  the Royal Society A: Mathematical, Physical and Engineering Sciences 379
  (2021) 20210021.

\bibitem{Lo2019}
Y.~S. Lo, M.~J. Borden, K.~Ravi-Chandar, C.~M. Landis, A phase-field model for
  fatigue crack growth, Journal of the Mechanics and Physics of Solids 132
  (2019) 103684.

\bibitem{Boldrini2016}
J.~L. Boldrini, E.~A.~B. de~Moraes, L.~R. Chiarelli, F.~G. Fumes, M.~L.
  Bittencourt, A non-isothermal thermodynamically consistent phase field
  framework for structural damage and fatigue, Computer Methods in Applied
  Mechanics and Engineering 312 (2016) 395--427.

\bibitem{Loew2020}
P.~J. Loew, B.~Peters, L.~A. Beex, Fatigue phase-field damage modeling of
  rubber using viscous dissipation: Crack nucleation and propagation, Mechanics
  of Materials 142 (2020) 103282.

\bibitem{Schreiber2020}
C.~Schreiber, C.~Kuhn, R.~Müller, T.~Zohdi, A phase field modeling approach of
  cyclic fatigue crack growth, International Journal of Fracture 225 (2020).

\bibitem{Alessi2018}
R.~Alessi, S.~Vidoli, L.~D. Lorenzis, A phenomenological approach to fatigue
  with a variational phase-field model: The one-dimensional case, Engineering
  Fracture Mechanics 190 (2018) 53--73.

\bibitem{Carrara2020}
P.~Carrara, M.~Ambati, R.~Alessi, L.~D. Lorenzis, A framework to model the
  fatigue behavior of brittle materials based on a variational phase-field
  approach, Computer Methods in Applied Mechanics and Engineering 361 (2020)
  112731.

\bibitem{Seiler2020}
M.~Seiler, T.~Linse, P.~Hantschke, M.~Kästner, An efficient phase-field model
  for fatigue fracture in ductile materials, Engineering Fracture Mechanics 224
  (2020).

\bibitem{Simoes2021}
M.~Simoes, E.~Martínez-Pañeda, Phase field modelling of fracture and fatigue
  in shape memory alloys, Computer Methods in Applied Mechanics and Engineering
  373 (2021) 113504.

\bibitem{Simoes2022}
M.~Simoes, C.~Braithwaite, A.~Makaya, E.~Martínez-Pañeda, Modelling fatigue
  crack growth in shape memory alloys, Fatigue \& Fracture of Engineering
  Materials \& Structures 45 (2022) 1243--1257.

\bibitem{Ai2022}
W.~Ai, B.~Wu, E.~Martínez-Pañeda, A coupled phase field formulation for
  modelling fatigue cracking in lithium-ion battery electrode particles,
  Journal of Power Sources 544 (2022) 231805.

\bibitem{Hasan2021}
M.~M. Hasan, T.~Baxevanis, A phase-field model for low-cycle fatigue of brittle
  materials, International Journal of Fatigue 150 (2021) 106297.

\bibitem{Golahmar2022}
A.~Golahmar, P.~K. Kristensen, C.~F. Niordson, E.~Martínez-Pañeda, A phase
  field model for hydrogen-assisted fatigue, International Journal of Fatigue
  154 (2022) 106521.

\bibitem{Seles2021}
K.~Seleš, F.~Aldakheel, Z.~Tonković, J.~Sorić, P.~Wriggers, A general
  phase-field model for fatigue failure in brittle and ductile solids,
  Computational Mechanics 67 (2021).

\bibitem{Ulloa2021}
J.~Ulloa, J.~Wambacq, R.~Alessi, G.~Degrande, S.~François, Phase-field
  modeling of fatigue coupled to cyclic plasticity in an energetic formulation,
  Computer Methods in Applied Mechanics and Engineering 373 (2021) 113473.

\bibitem{Khalil2022}
Z.~Khalil, A.~Y. Elghazouli, E.~Martínez-Pañeda, A generalised phase field
  model for fatigue crack growth in elastic–plastic solids with an efficient
  monolithic solver, Computer Methods in Applied Mechanics and Engineering 388
  (2022) 114286.

\bibitem{Pham2011}
K.~Pham, H.~Amor, J.~J. Marigo, C.~Maurini, Gradient damage models and their
  use to approximate brittle fracture, International Journal of Damage
  Mechanics 20 (2011) 618--652.

\bibitem{Wu2020c}
J.-Y. Wu, Y.~Huang, V.~P. Nguyen, On the bfgs monolithic algorithm for the
  unified phase field damage theory, Computer Methods in Applied Mechanics and
  Engineering 360 (2020) 112704.

\bibitem{Kristensen2020}
P.~K. Kristensen, E.~Martínez-Pañeda, Phase field fracture modelling using
  quasi-newton methods and a new adaptive step scheme, Theoretical and Applied
  Fracture Mechanics 107 (2020) 102446.

\bibitem{Provatas2011}
N.~Provatas, K.~Elder, Phase-Field Methods in Materials Science and
  Engineering, John Wiley \& Sons, 2011.

\bibitem{Cui2021}
C.~Cui, R.~Ma, E.~Martínez-Pañeda, A phase field formulation for
  dissolution-driven stress corrosion cracking, Journal of the Mechanics and
  Physics of Solids 147 (2021) 104254.

\bibitem{Ambati2015a}
M.~Ambati, T.~Gerasimov, L.~D. Lorenzis, A review on phase-field models of
  brittle fracture and a new fast hybrid formulation, Computational Mechanics
  55 (2015) 383--405.

\bibitem{Miehe2010a}
C.~Miehe, F.~Welshinger, M.~Hofacker, Thermodynamically consistent phase-field
  models of fracture: Variational principles and multi-field fe
  implementations, International Journal for Numerical Methods in Engineering
  83 (2010) 1273--1311.

\bibitem{Freddi2010}
F.~Freddi, G.~Royer-Carfagni, Regularized variational theories of fracture: A
  unified approach, Journal of the Mechanics and Physics of Solids 58 (2010)
  1154--1174.

\bibitem{Amor2009}
H.~Amor, J.~J. Marigo, C.~Maurini, Regularized formulation of the variational
  brittle fracture with unilateral contact: Numerical experiments, Journal of
  the Mechanics and Physics of Solids 57 (2009) 1209--1229.

\bibitem{Miehe2010b}
C.~Miehe, M.~Hofacker, F.~Welschinger, A phase field model for rate-independent
  crack propagation: Robust algorithmic implementation based on operator
  splits, Computer Methods in Applied Mechanics and Engineering 199 (2010)
  2765--2778.

\bibitem{Tanne2018}
E.~Tanné, T.~Li, B.~Bourdin, J.-J. Marigo, C.~Maurini, Crack nucleation in
  variational phase-field models of brittle fracture, Journal of the Mechanics
  and Physics of Solids 110 (2018) 80--99.

\bibitem{Walker1970}
K.~Walker, The effect of stress ratio during crack propagation and fatigue for
  2024-t3 and 7075-t6 aluminum, effects of environment and complex load history
  on fatigue life, ASTM STP 462 (1970).

\bibitem{Smith1970}
K.~N. Smith, P.~Watson, T.~H. Topper, Stress- strain function for the fatigue
  of metals, Journal of Material, ASTM 5 (1970) 767--778.

\bibitem{MIL1998}
MIL-HDBK-5H, Military Handbook: Metallic Materials and Elements for Aerospace
  Vehicle Structures, U.S. Department of Defense, 1998.

\bibitem{Dowling2009}
N.~E. Dowling, C.~A. Calhoun, A.~Arcari, Mean stress effects in stress-life
  fatigue and the walker equation, Fatigue and Fracture of Engineering
  Materials and Structures 32 (2009) 163--179.

\bibitem{McClung1991}
R.~C. McClung, {Crack closure and plastic zone sizes in fatigue}, Fatigue {\&}
  Fracture of Engineering Materials {\&} Structures 14~(4) (1991) 455--468.

\end{thebibliography}

\small
\bibliographystyle{elsarticle-num}

\end{document}